\documentclass[prb,twocolumn,preprintnumbers,amsmath,amssymb,showpacs,superscriptaddress,floatfix]{revtex4}

\usepackage{graphicx}
\usepackage{dcolumn}
\usepackage{bm}

\begin{document}


\title{Resonant soft x-ray scattering from stepped surfaces of SrTiO$_{3}$}

\author{J. Schlappa}
\altaffiliation[Present adress: ]{Swiss Light Source, Paul Scherrer Institut, Villingen, Switzerland}
\affiliation{{II.} Physikalisches Institut, Universit{\"a}t zu
K{\"o}ln, Z{\"u}lpicher Str.~77, D-50937 K{\"o}ln, Germany}
\author{C. Sch{\"u}{\ss}ler-Langeheine}
\email [Corresponding author: ]{schuessler@ph2.uni-koeln.de}
\affiliation{{II.} Physikalisches Institut, Universit{\"a}t zu K{\"o}ln,
Z{\"u}lpicher Str.~77, D-50937 K{\"o}ln, Germany}
\author{C.~F. Chang}
\affiliation{{II.} Physikalisches Institut, Universit{\"a}t zu K{\"o}ln,
Z{\"u}lpicher Str.~77, D-50937 K{\"o}ln, Germany}
\author{Z. Hu}
\affiliation{{II.} Physikalisches Institut, Universit{\"a}t zu K{\"o}ln,
Z{\"u}lpicher Str.~77, D-50937 K{\"o}ln, Germany}
\author{E. Schierle}
\altaffiliation[Present adress: ]{HMI c/o BESSY, Berlin, Germany}
\affiliation{Institut f{\"ur} Experimentalphysik, Freie Universit{\"a}t Berlin,
Arnimallee 14, D-14195 Berlin, Germany}
\author{H. Ott}
\altaffiliation[Present adress: ]{Shell International
Exploration \& Production B.V., Rijswijk,
The Netherlands}
\affiliation{Institut f{\"ur} Experimentalphysik, Freie Universit{\"a}t Berlin,
Arnimallee 14, D-14195 Berlin, Germany}
\author{E. Weschke}
\altaffiliation[Present adress: ]{HMI c/o BESSY, Berlin, Germany}
\affiliation{Institut f{\"ur} Experimentalphysik, Freie Universit{\"a}t Berlin,
Arnimallee 14, D-14195 Berlin, Germany}
\author{G. Kaindl}
\affiliation{Institut f{\"ur} Experimentalphysik, Freie Universit{\"a}t Berlin,
Arnimallee 14, D-14195 Berlin, Germany}
\author{M. Huijben}
\affiliation{Faculty of Science and Technology and MESA+ Institute for Nanotechnology, University of Twente, Enschede, The Netherlands}%
\author{G. Rijnders}
\affiliation{Faculty of Science and Technology and MESA+ Institute for Nanotechnology, University of Twente, Enschede, The Netherlands}%
\author{D. H. A. Blank}
\affiliation{Faculty of Science and Technology and MESA+ Institute for Nanotechnology, University of Twente, Enschede, The Netherlands}%
\author{L. H. Tjeng}
\affiliation{{II.} Physikalisches Institut, Universit{\"a}t zu K{\"o}ln,
Z{\"u}lpicher Str.~77, D-50937 K{\"o}ln, Germany}%

\date{\today}

\begin{abstract}
We studied the resonant diffraction signal from stepped surfaces of SrTiO$_3$ at the Ti $2p \rightarrow 3d$ ($L_{2,3}$) resonance in comparison with x-ray absorption (XAS) and specular reflectivity data. The steps on the surface form an artificial superstructure suited as a model system for resonant soft x-ray diffraction. A small step density on the surface is sufficient to produce a well defined diffraction peak, showing the high sensitivity of the method. At larger incidence angles, the resonant diffraction spectrum from the steps on the surface resembles the spectrum for specular reflectivity. Both deviate from the XAS data in the relative peak intensities and positions of the peak maxima. We determined the optical parameters of the sample across the resonance and found that the differences between the XAS and scattering spectra reflect the different quantities probed in the different signals. When recorded at low incidence or detection angles, XAS and specular reflectivity spectra are distorted by the changes of the angle of total reflection with energy. Also the step peak spectra, though less affected, show an energy shift of the peak maxima in grazing incidence geometry.
\end{abstract}


\maketitle

\section{Introduction}

Resonant x-ray diffraction in the soft x-ray range (RSXD) combines a high spectroscopic sensitivity with momentum resolution. While its development started only recently, it promises to be a powerful tool for the study of phenomena like charge and orbital order, as they are found in many correlated electron systems. \cite{Imada1998} Corresponding RSXD experimental results have been published in the last year from  manganese systems,\cite{Wilkins2003_1,Wilkins2003_2,Dhesi2004,Thomas2004,Wilkins2005,Staub2005,wilkins:06a,staub:06a,smadici:07b,grenier:07a,grenier:07b,bodenthin:08a,garcia:08a} magnetite,\cite{Huang2006,schlappa:08a} nickelates,\cite{ourPRL2005,Scagnoli2006,staub:07a} cuprates,\cite{Abbamonte2002,Abbamonte2004,Abbamonte2005,Rusydi2006,rusydi:07a,smadici:07a,rusydi:08a} and ruthenates.\cite{Zegkinoglou2005} It is a bit unfortunate, though, that the development of this new technique has so far been essentially done using only these interesting but complex materials. This is a consequence of the rather long photon wavelengths in the soft x-ray range. A RSXD experiment requires a sample with periodicities in the nm range, which is not easy to find among simple systems. Using complex materials instead, one has to solve two things at the same time: the development of the new technique and the complex physics of the material itself.

Here we present a suited simple model system for resonant soft x-ray scattering techniques: stepped surfaces of single crystalline SrTiO$_3$. The system is electronically and structurally simple: The structure is cubic perovskite and because of the empty $3d$ shell of the Ti ions no electron-correlation effects need to be considered. In order to match the rather long photon wavelengths at the Ti $2p \rightarrow 3d$ ($L_{2,3}$) resonance, we used the steps on vicinal surfaces as artificial superstructures with period lengths between about 20 nm and 70 nm. Samples with such surfaces allow to study the diffraction signal caused by the steps, the specular reflectivity, and the x-ray absorption spectroscopy (XAS) signal and to compare these. One question that can be addressed with such a model is how sensitive RSXD is, i.e., what concentration of scatterers can be detected. Unlike in the conventional x-ray range, the optical parameters change strongly across a resonance in the soft x-ray range such that refraction effects may matter. For the model system we could study, how these x-ray optical effects affect the observed resonance data. 
  
The rest of this contribution is organized as follows: Sect.~II describes the sample system, Sect.~III the experimental setup, Sect.~IV the experimental results from a stepped surface at large incidence and diffraction angles, Sect.~V the optical parameters of Ti in SrTiO$_3$ and Sect.~VI how the different signals are affected by x-ray optical effects at different angles. In Sect.~VII the results are summarized.

\section{S\lowercase{r}T\lowercase{i}O$_{3}$}
\label{srtio3_1}

SrTiO$_{3}$ is a particularly appropriate system for model studies, because of its simple electronic  and crystalline structure. The formal valence of the ions in this compound is Sr$^{2+}$ $(5s^0)$, Ti$^{4+}$ $(3d^0)$ and O$^{2-}$ ($2p^6$), meaning that the electron shells of all ions are
either fully occupied or empty in the ground state. It is a band insulator and possesses no local magnetic moments. The crystal structure at room temperature is cubic perovskite (space group $Pm3m$), with a lattice constant of 3.905~\AA. The Ti-ions are therefore embedded in an octahedral crystal field, generated by the neighboring oxygen ions. Due to this symmetry, the optical properties of the Ti-ions are isotropic, which means independent on the polarization direction of the scattered light.

\begin{figure}[t]
\begin{center}
\includegraphics[clip,width=0.45\textwidth]{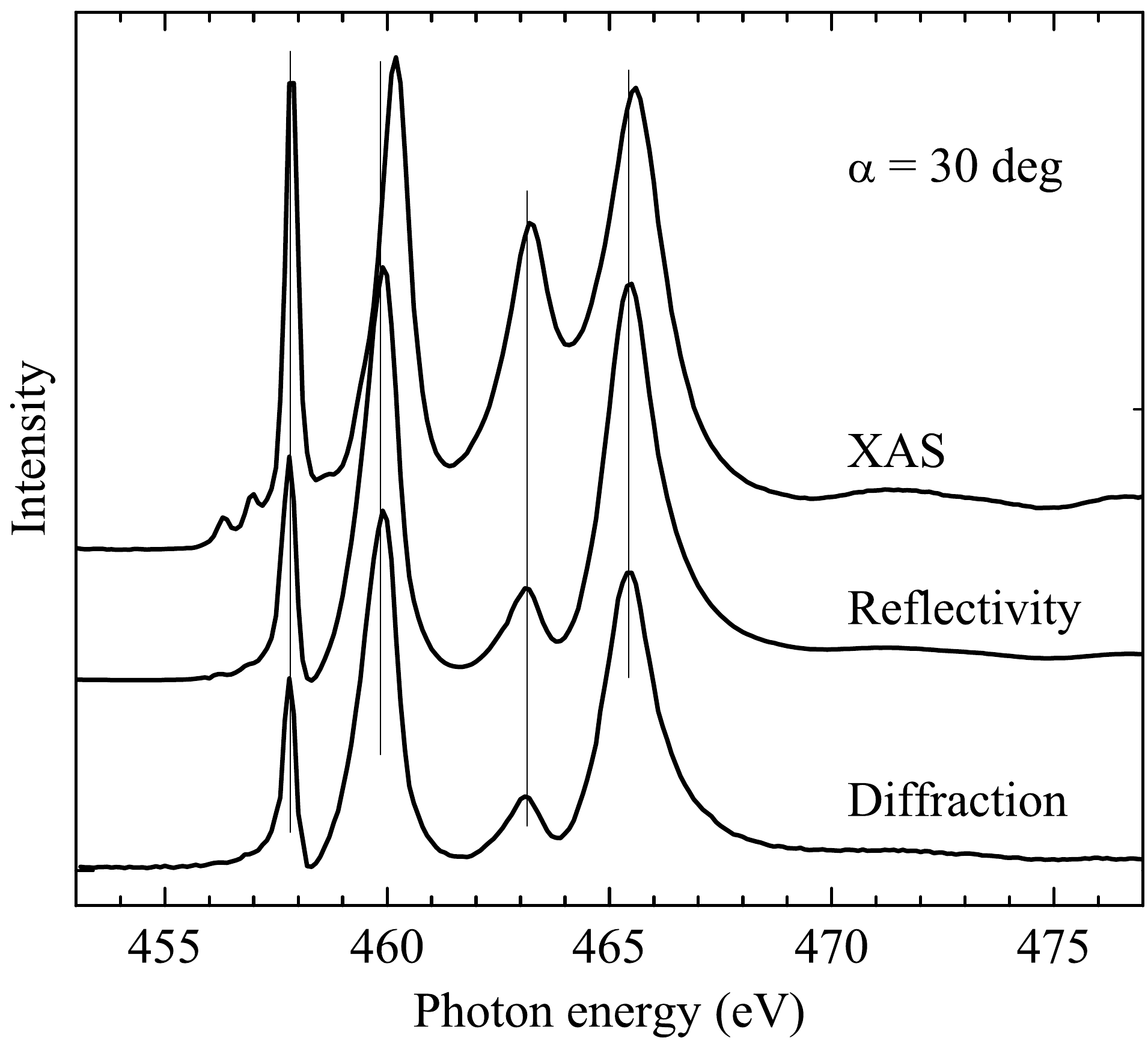}
\caption{Energy dependence of the x-ray absorption (XAS), reflectivity
and diffraction signal across Ti $L_{2,3}$ edge for a 0.3~deg-inclined
SrTiO$_3$ sample. The vertical lines denote the positions of the major peaks in the diffraction spectrum}\label{XRS30}
\end{center}
\end{figure}

At energies around the Ti $L_{2,3}$ edge ($\sim 460$~eV) the optical properties of the compound are totally dominated by the dipole allowed Ti $2p \rightarrow 3d$ transition. The Ti$^{4+}$ ions are excited from the $2p^63d^0$ to a $2p^53d^1$ multiplet state, under the formation of a bound excitonic state between the core hole and the excited electron. Therefore this process can be well described in terms of a local picture, assuming that the electron is localized at the TiO$_6$ cluster. The absorption spectra of $d^{0}$ systems in $O_h$ symmetry at transition metal $L_{2,3}$ edges have been extensively studied and modeled by de Groot et al.,\cite{deGroot1990} applying atomic multiplet calculations including crystal-field interactions. The spectra can be fully described by the $2p^53d^1$ multiplet, which consists of seven visible lines (see XAS curve in Fig.~\ref{XRS30}). The four most intense are split into two groups, due to the spin-orbit coupling of the core hole, such that the two at lower energies belong to the $L_3$ edge, whereas the other two to the $L_2$ edge. The splitting within each group is mainly determined by the crystal field-splitting of the $O_h$ symmetry, $10Dq$. Two of the weaker lines are on the low energy side of the spectrum, the third in between the two $L_3$ main lines. They are due to the Coulomb and exchange interactions within the multiplet.

\section{Experimental}
\label{srtio3_2}

\begin{figure}[t]
\begin{center}
\includegraphics[clip,width=0.45\textwidth]{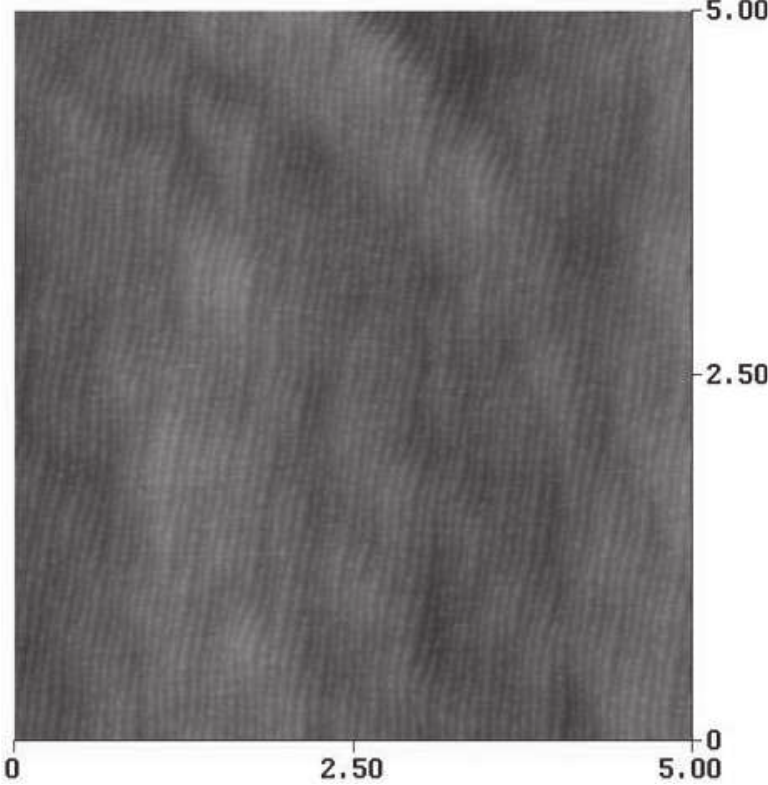}
\caption{Atomic force microscopy (AFM) image from one of the used stepped SrTiO$_3$ surfaces with a miscut angle of $\approx$ 0.3$^\circ$. The field of vision is $5 \times 5$ $\mu$m$^2$.}\label{AFM}
\end{center}
\end{figure}

Samples with stepped surfaces were prepared at the University of Twente, The Netherlands. In order to obtain a well defined TiO$_2$-terminated surface, the SrTiO$_3$ single crystals were etched in buffered HF and subsequently annealed in oxygen.\cite{Koster1998} The samples had the form of square plates with an edge length of 10~mm and a thickness of 0.5 or 1~mm. The surface orientation was essentially (001). For the diffraction experiments we used samples with miscut angles, i.e., the difference between the averaged surface and the surface of one terrace on the surface, of about 0.3 and about 1 degrees. This corresponds to a terrace width of 70 and 20~nm, respectively. The samples were characterized by atomic-force microscopy (AFM) and x-ray absorption spectroscopy (XAS). Fig. \ref{AFM} shows an AFM picture of one of the samples. The steps are equally spaced, though slightly wavy, leading to a regular stripe-like surface structure. The edges of the terraces are almost parallel to the sample edge, which has the direction [100].

For the scattering experiment the stepped samples were mounted such, that the edge of the steps was oriented perpendicular to the scattering plane and the incident beam was pointing toward the steps (see Fig. \ref{StepsCart}). After transfer into UHV, the samples were annealed to 300 $^\circ$C in an oxygen atmosphere of $5\cdot10^{-5}$~mbar for 30~min, to desorb gas particles adhering to the surface.

The experiments were performed at the UE52-SGM1 beamline at BESSY using the UHV soft x-ray diffractometer designed at the FU Berlin. The scattering geometry was horizontal with the light linearly polarized parallel to the scattering plane ($\pi$-direction). The photons were detected using a photo-diode mounted behind a rectangular slit. For the energy scans from the step peaks the detector position and the photon energy was scanned simultaneously in order to stay on the diffraction peak. The detector acceptance was chosen such that the whole peak is detected. The energy dependence of the background below the peak was measured separately with the detector angles shifted by 3 degrees and was subtracted from the peak spectra. The x-ray absorption spectra were obtained from the total electron yield signal recorded with a channeltron electron detector.

\section{Diffraction data}
\label{srtio3_3}

\begin{figure}[t]
\begin{center}
\includegraphics[clip,width=0.45\textwidth]{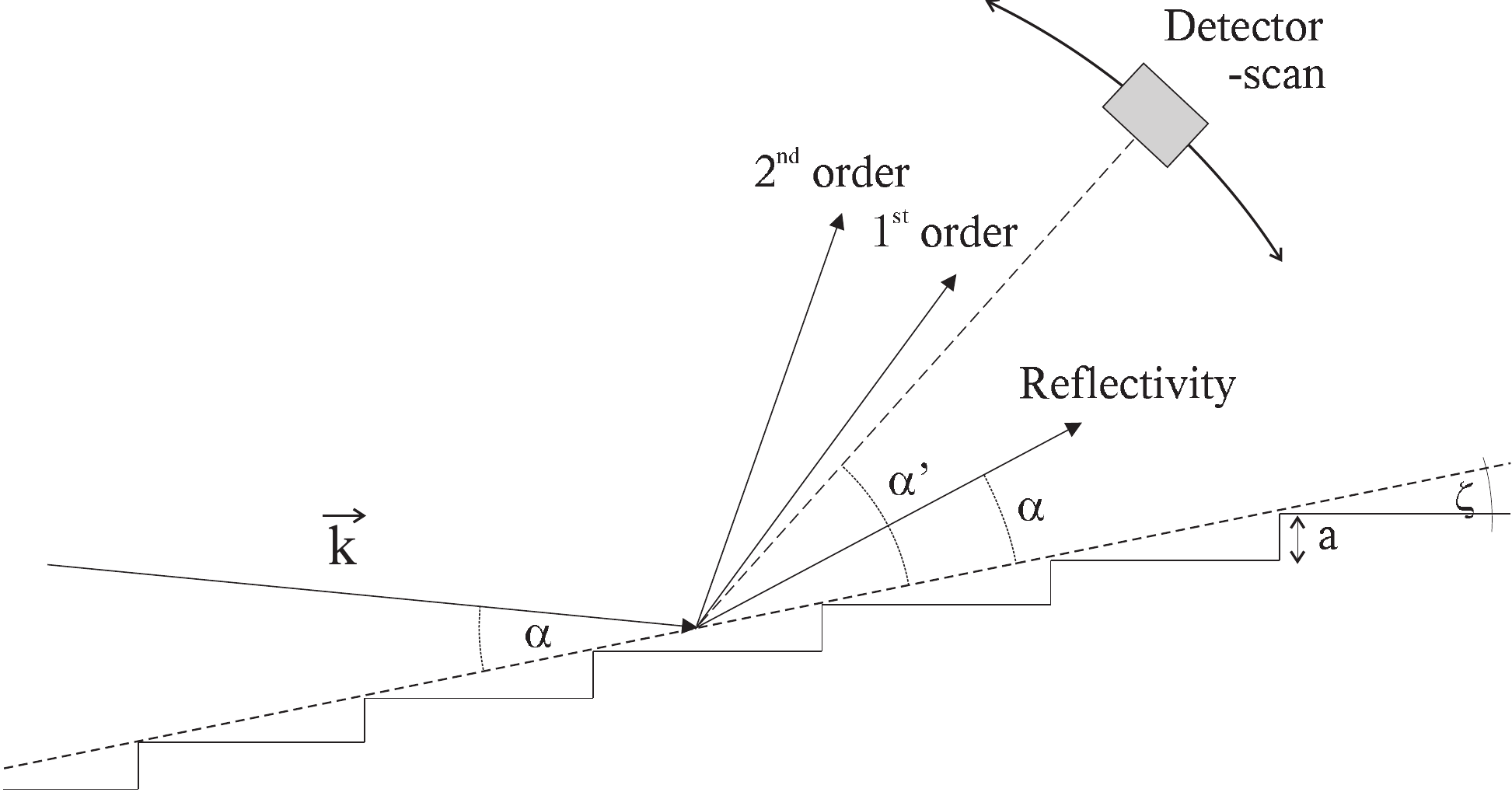}
\caption{Schematics of the scattering geometry for diffraction from a stepped
surface. $\vec{k}$ denotes the wave vector of the incoming photon, $\alpha$ the angle between $\vec{k}$ and the averaged sample surface, which is at an angle $\zeta$ with the surface of one of the terraces, $\alpha'$ the angle between the averaged surface and the detector and $a$ the lattice constant of SrTiO$_3$.}\label{StepsCart}
\end{center}
\end{figure}

A regular arrangement of steps on a surface has the properties of a reflection grating. When the height of these steps is equal to the lattice parameter $a$ and $\zeta$ is the miscut angle, the width of the terraces will be equal to $a/\tan\zeta$. A diffraction signal can be observed when the following condition is fulfilled
\begin{equation}
    \frac{a}{\tan\zeta} (\cos\alpha - \cos\alpha ') = m \lambda
\label{eq:grating}
\end{equation}
 with the integer $m$ giving the diffraction oder. The angles of incidence and diffraction, $\alpha$ and $\alpha '$, are expressed here with respect to the averaged surface normal (Fig.~\ref{StepsCart}), $\lambda$ is the x-ray wavelength, which is for the Ti $L_{2,3}$ resonance about 27 {\AA}.
For a 2-dimensional superstructure the reciprocal lattice consists of 1-dimensional rods, rather than of a 3-dimensional array of points. If the wavelength is shorter than the terrace width, the diffraction condition of Eq.~\ref{eq:grating} can be fulfilled for any chosen angle of incidence. For a chosen angle of incidence several reflections can be observed, as illustrated in Fig.~\ref{StepsCart}, one belonging to the specular reflectivity ($m=0$, $\alpha'=\alpha$) and others to diffraction from the terraces. These reflections can be probed, e.g. by scanning the detector angle $\alpha '$.

\begin{figure}[t]
\begin{center}
\includegraphics[clip,width=0.45\textwidth]{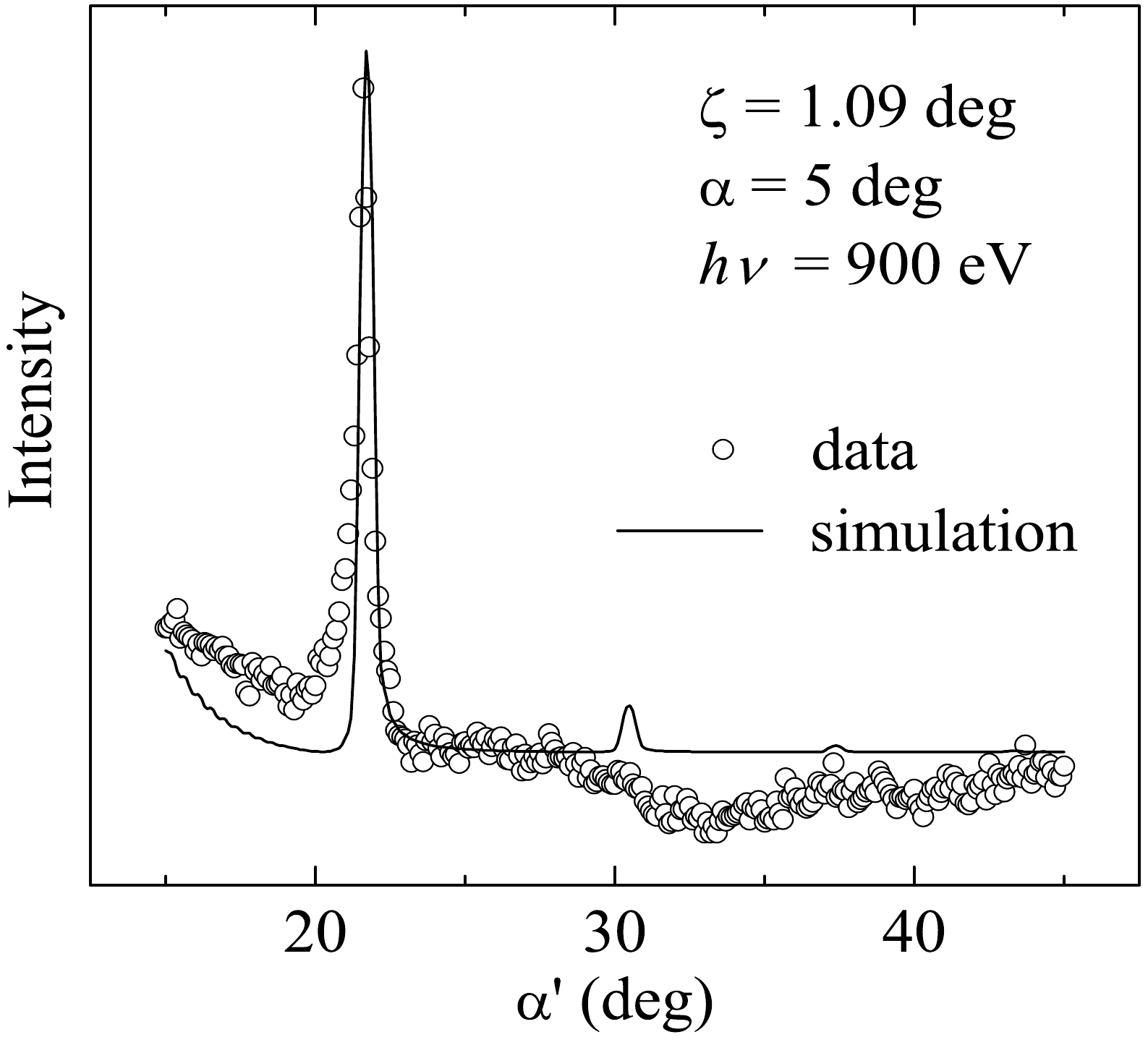}
\caption{Detector-angle scan across the diffraction peaks from a stepped 1.09~deg-inclined SrTiO$_3$ sample, recorded off resonance.}\label{Bessyrep}
\end{center}
\end{figure}

In order to test the signal strength, we first performed an experiment far from resonance. Fig. \ref{Bessyrep} shows an off-resonance scan measured from 1.09~degrees-inclined sample by changing the detector angle. The angle of incidence was kept at 5 degrees and the photon energy was 900~eV, lying far above the Ti resonance. The symbols in the diagram represent experimental data whereas the solid line is a simulation using a simple kinematical model of scattering from steps. The data shows a pronounced maximum around 21.7$^\circ$, which is the position for the first order reflection. The observation of this signal clearly demonstrates the suitability of this model system for soft x-ray diffraction. Off the Ti resonance and for the chosen scattering geometry we can estimate the volume of Ti ions at step edges to about $5 \cdot 10^{-5}$ of the total scattering volume. The clear signal observed even off resonance shows that diffraction using soft x-ray energies is sensitive enough to study even very ``dilute'' systems.

The second order peak cannot be resolved from the background signal. Apparently its intensity is lower than expected for an ideal system, probably due to the waviness of the step edges. The angular width of the measured 1st order diffraction peak shows that the widths of the terraces on the surface vary around the average value. In the simulation a Gaussian distribution of widths around $a/\tan\zeta$ was considered. The increase of background toward low detector angles is due to the tail of the reflectivity.

When the photon energy is tuned to the Ti $L_{2}$- or $L_{3}$-edge, the diffraction signal increases considerably by about one order of magnitude. The intensity of the diffraction signal varies strongly across the resonance edge, as energy is changed. Fig. \ref{XRS30} shows an energy scan of the diffraction signal (lower curve), compared with the x-ray absorption (XAS) and reflectivity spectrum for the same energy range. All spectra were taken from a 0.31~deg-inclined sample at 30~deg incidence angle. For the step peak the detector was at 34 degrees. The diffraction and reflectivity spectra look almost identical. They show the four main resonance peaks discussed above, which are also present in the absorption spectrum. The relative intensities of the main peaks in the absorption spectrum and the scattering data are different with the weak lines more pronounced in the XAS data than in the diffraction and reflectivity data. In fact it is not expected for the absorption spectrum to look like the diffraction or reflectivity signal since XAS is probing only the imaginary part of the scattering amplitude, whereas the other two signals are determined by the squared modulus of the sum of the real and imaginary part of the scattering amplitude.

For a single harmonic oscillator the absorption spectrum is a Lorentzian \cite{als-nielsen2001} of width $\Gamma$ and intensity $A$. The scattered intensity determined by the norm squared of the real and the imaginary part is (ignoring the interference with non-resonant contributions to the scattering amplitude) again a Lorentzian of the same width but with an intensity proportional to $A^2/\Gamma$. This leads to a general enhancement of strong and narrow resonance lines in the scattered signal. For the case of several overlapping resonances, as we have them here, also the interference between the different oscillators matters. In fact, in the reflectivity and diffraction spectra, the peak maxima of the major peaks except for that one at lowest photon energies are shifted by up to 0.3 eV toward lower energies with respect to the XAS data. In order to check whether the observed energy shift is indeed an interference effect, we determined the optical parameters of SrTiO$_3$. 

\section{Optical parameters}
\label{srtio3_4}

Off resonance the index of refraction is tabulated, \cite{Henke1993} but these tables are not accurate enough near resonances. We therefore measured the specular reflectivity from a flat sample of SrTiO$_3$ as a function of photon energy around the Ti $L_{2,3}$ resonance using $\pi$-polarized light. 

The intensity of the reflectivity signal is described by Fresnel equations, which are functions of the angle of incidence and the refraction index.\cite{Jackson} For the interface between vacuum and a medium with refractive index $n$ the amplitude of the reflected wave for $\pi$ polarization, $E_\pi$, is 
\begin{equation}
E_\pi (\alpha,n) = E_0 \cdot \frac{n^2 \sin \alpha - \sqrt{n^2 - \cos^2 \alpha}}
                                  {n^2 \sin \alpha + \sqrt{n^2 - \cos^2 \alpha}}
\label{eq:Fresnel}
\end{equation}
with $E_0$ being the intensity of the incident wave.
Therefore, knowing the dependence of the signal on the angle of incidence $\alpha$, the index of refraction $n$ can be obtained experimentally and consequently, according to $n = 1 - \delta + i\beta$, also the optical constants $\delta$ and $\beta$.

\begin{figure}[t]
\begin{center}
\includegraphics[width=0.45\textwidth]{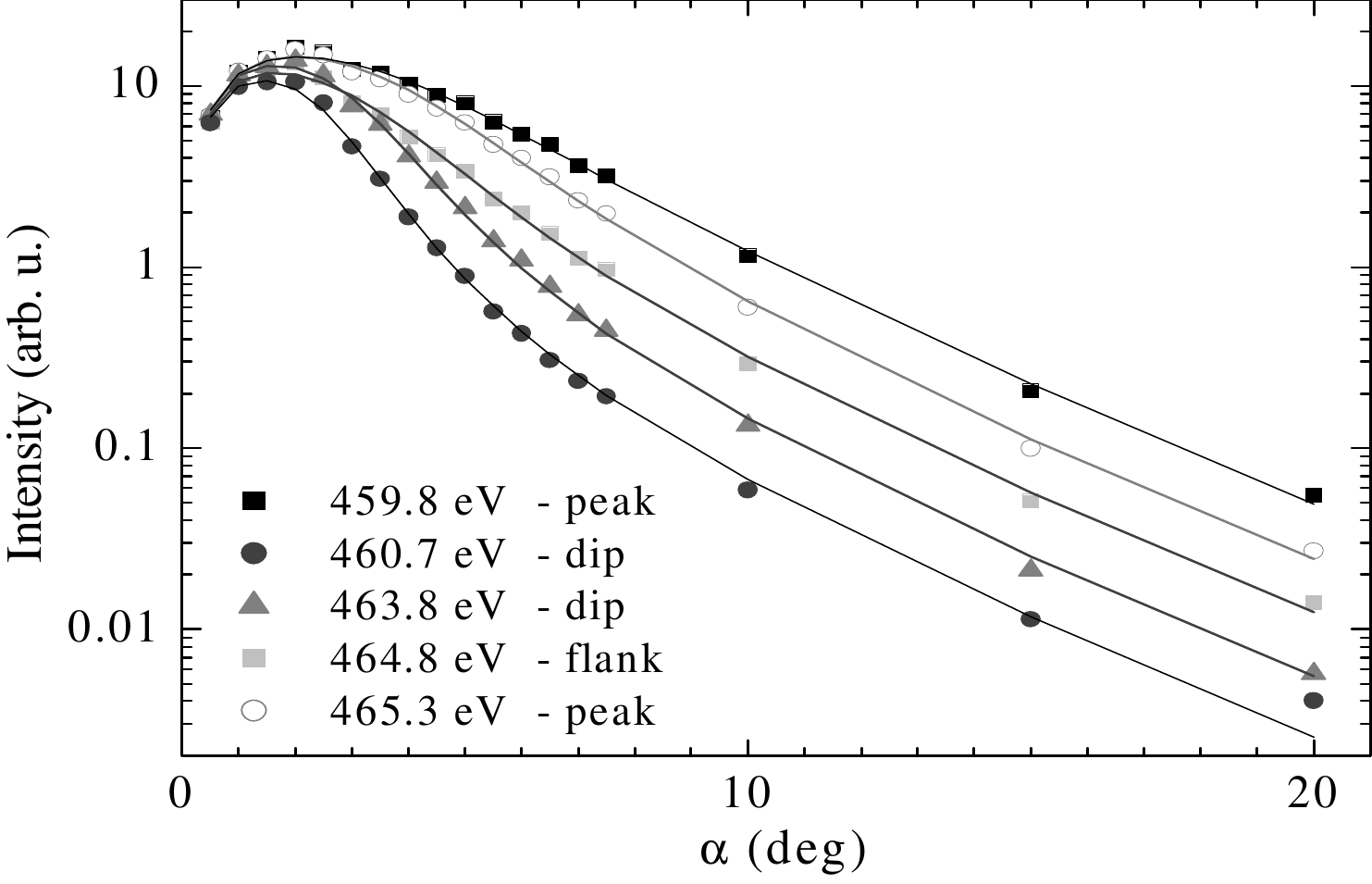}
\caption{Reflectivity curves (symbols) as a function of the angle of incidence
$\alpha$ for different characteristic photon energies on a logarithmic scale; the solid lines are fits applying the Fresnel equation and the footprint effect at low incidence angles. }\label{Reflectiv}
\end{center}
\end{figure}

Fig. \ref{Reflectiv} shows reflectivity data for selected energies (symbols), together with the result of a least square fit (lines). The intensity of the signal decreases approximately exponentially with larger angles of incidence. At low angles total external reflection occurs leading to the maximum around 2 degrees. The decrease of signal for very small incidence angles seen in the data is due to the fact that for very small incidence angles the footprint of the x-ray beam on the sample surface becomes larger than the sample itself, leading to a loss of intensity. In this region the intensity of the light falling on the sample is proportional to $(x\sin\alpha)/b$, where $x$ is the sample width and $b$ the beam-width in the scattering plane.

\begin{figure}[t]
\begin{center}
\includegraphics[width=0.45\textwidth]{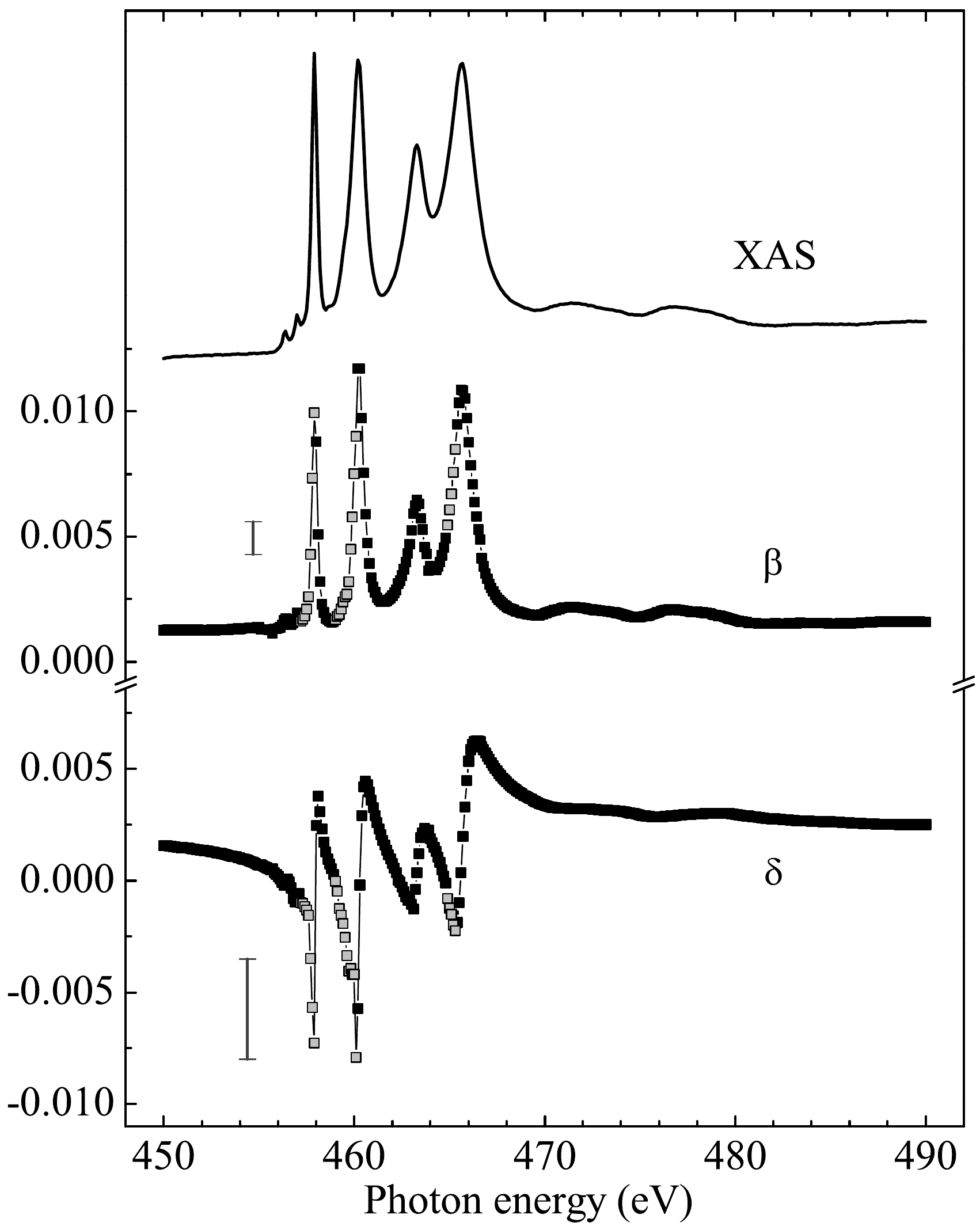}
\caption{Optical parameters $\delta$ and $\beta$ together with the
absorption spectrum (XAS) at 30~deg of incidence. The error bars
indicate the uncertainty for the gray marked points. For the black marked points the error margins are around 10 percent.}\label{DeltaBeta}
\end{center}
\end{figure}

The data were analyzed using the Fresnel equation for $\pi$-polarized light (\ref{eq:Fresnel}) and a Debye-Waller like damping to account for the sample roughness of 3~\AA. For large negative values of $\delta$ (indicated in Fig.~\ref{DeltaBeta} by the gray points) the shape of the reflectivity curve depends only very weakly on the value of $n$ for a wide range of ($\delta,\beta$) sets.\cite{Soufli1997} The fit does hence not converge in this region. The XAS data, however, provide a second independent information about $\beta$ via the relation $\beta = - \lambda \mu / 4 \pi$,\cite{als-nielsen2001} with $\mu$ being the linear absorption coefficient probed in an XAS experiment. This provides a constraint for the value of $\beta$ in the fit, which was used to determine $\delta$ in these critical regions. 

The thus obtained optical constants $\delta$ and $\beta$ are shown in Fig.~\ref{DeltaBeta}, together with the absorption spectrum (XAS) at 30~deg incidence angle (upper curve). The error bars denote the maximum uncertainty for the data points denoted by grey symbols. For values shown with black symbols the uncertainty for $\delta$ amounts to 10 percent, whereas for $\beta$ it is a little bit smaller.
Above the resonance at 490~eV, $\delta$ is approximately $2.5\cdot 10^{-3}$ and $\beta$ is $1.6\cdot 10^{-3}$, which agrees very well with the tabulated x-ray data, \cite{Henke1993}
giving for this energy $2.3\cdot 10^{-3}$ and $1.5\cdot 10^{-3}$, respectively (using a mass density, $\rho_a$ of $5$~g/cm$^3$ for the sample).

Across the resonance the optical constants are strongly energy-dependent. While $\beta$ resembles the absorption spectrum, reaching peak values at each resonance - with the maximum of $12\cdot 10^{-3}$ at 460.25~eV, $\delta$ behaves similar to the negative derivative of $\beta$. $\delta$ changes sign each time a strong absorption peak is crossed; coming from the low-energy side it decreases steadily reaching a negative-signed minimum just before the absorption peak; directly above the absorption maximum it increases steeply toward a positive-signed maximum, crossing the zero line exactly at the resonance position, and decreases again more slowly afterwards, crossing the zero line in the vicinity of a local $\beta$ minimum. The largest absolute value that $\delta$ reaches is negative-signed and amounts to $-8\cdot 10^{-3}$ at 460.1~eV.

\begin{figure}[!t]
\begin{center}
\includegraphics[width=0.45\textwidth]{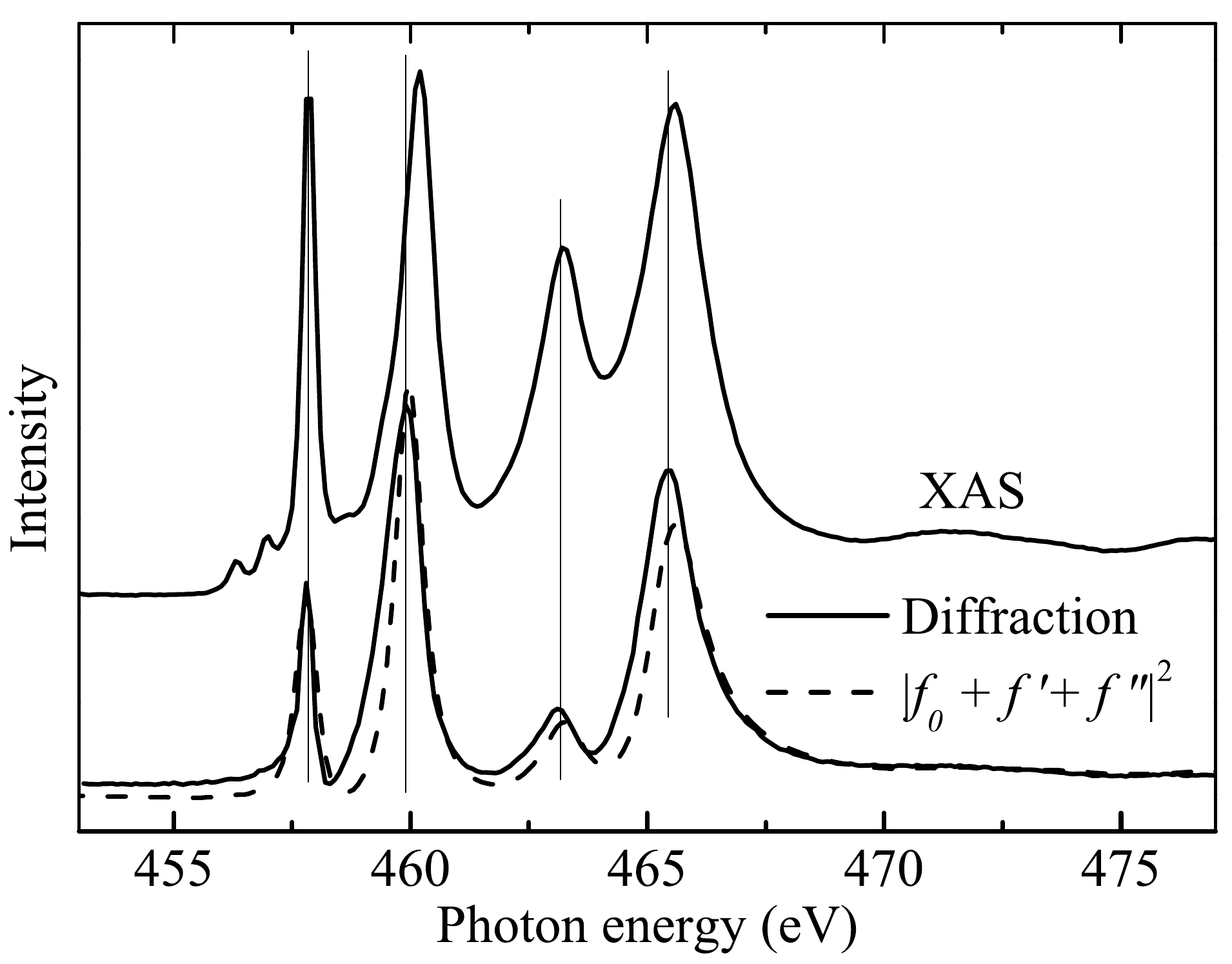}
\caption{XAS and diffraction data from Fig.~\ref{XRS30} with the result of a simulation of the diffraction spectrum using the optical parameters from Fig.~\ref{DeltaBeta} (dashed line).}\label{simresults}
\end{center}
\end{figure}

The extreme variation of $\delta$ toward negative values before each resonance strongly dominates the optical properties of the material, changing its optical density. It is remarkable that for certain energies the substance becomes optically more dense than vacuum (each time $\delta$ is negative), whereas for the conventional x-ray range (photon energies $\geq 10$~keV) it is known that matter is always optically less dense than vacuum. This fact demonstrates that the oscillator or atomic scattering strength is extremely large in the vicinity of the soft x-ray absorption edges. 

\section{Optical effects in the spectra}
\label{srtio3_4a}

\begin{figure*}[!t]
\begin{center}
\includegraphics[width=0.9\textwidth]{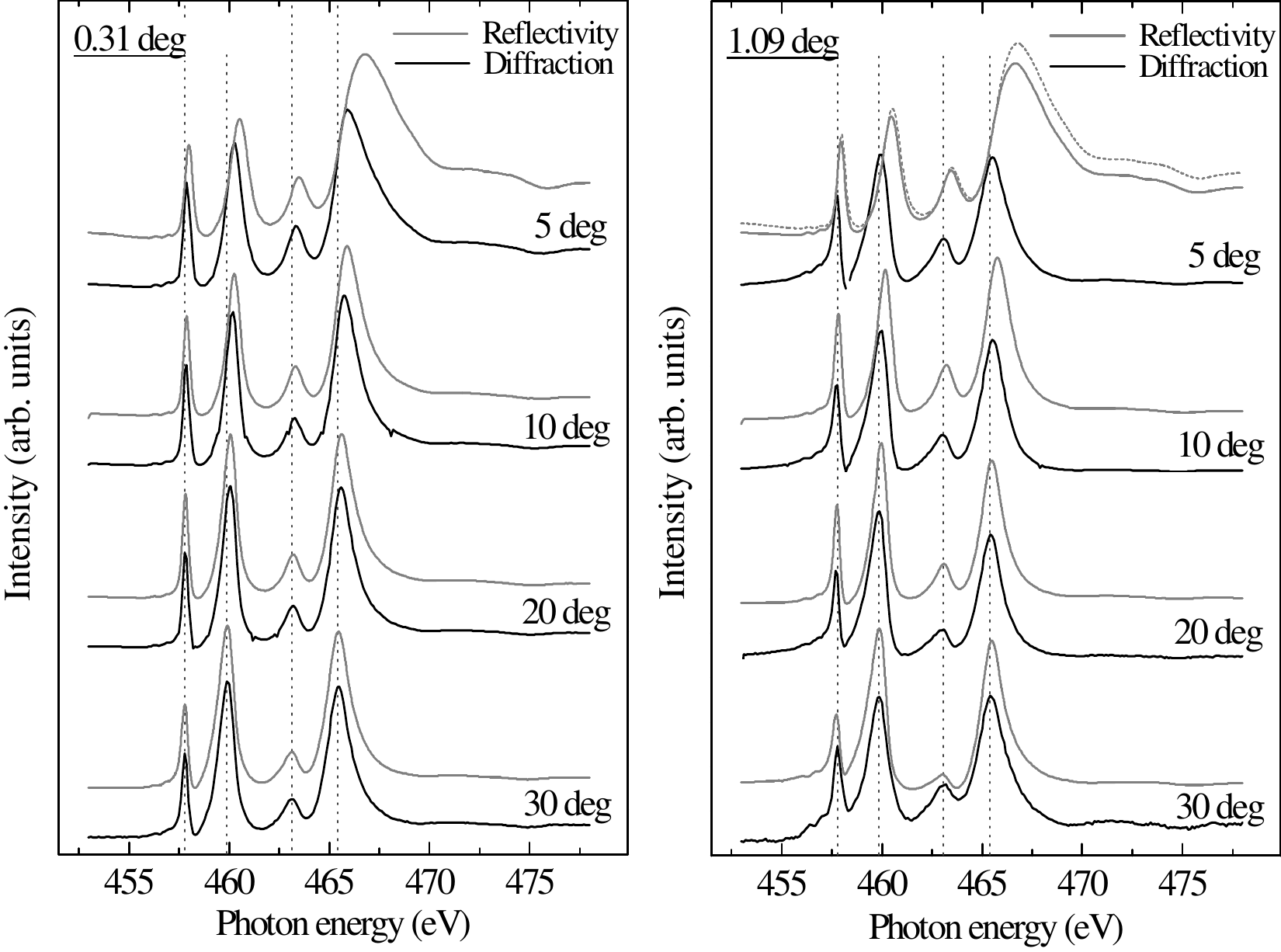}
\caption{Reflectivity and diffraction data from 0.31~deg (left)
and 1.09~deg (right) inclined samples for different incidence
angles ($\alpha$ = 5°, 10°, 20°, 30°). The dotted curve for
$\alpha = 5°$ at the right side is
reflectivity of the 0.31~deg sample plotted for comparison.}\label{data}
\end{center}
\end{figure*}

In Fig.~\ref{simresults} we present the result of a simulation of the diffraction-peak spectrum from Fig.~\ref{XRS30} using the optical parameters. The dashed curve was calculated from
\begin{equation}
 I = \left| f_0+f'+if'' \right|^2 = \left| \frac{2 \pi}{\lambda^2 \rho_a r_0} (\delta - i \beta) \right|^2
\end{equation}
($r_0 \approx 2.82 \cdot 10^{-5}$ {\AA} is the classical electron radius) and was broadened by 300 meV in order to account for the different energy resolutions in the respective experiments. The simulated curve describes the experimental curve fairly well including the relative intensities. In particular the shifted energy position of the second sharp maximum is essentially reproduced. Even though the agreement is less satisfactory for the $L_2$ part of the spectrum, this simple purely optical simulation indicates that the observed spectral differences between XAS and scattering data are mainly the result of the different optical quantities probed in the different experiments.

While the data from Fig.~\ref{XRS30} were recorded at large incidence and detection angles of 30 and 34 degrees, at small angles the effect of refraction of the photons in the sample becomes important and the consequent appearance of total reflection for certain energies across resonances.

How these changes affect the spectra is demonstrated in Fig.~\ref{data} where the corresponding signals are shown for two different surfaces and for different incidence angles $\alpha$ = 5, 10, 20 and 30 degrees. The corresponding detection angles $\alpha'$ varied around 16.5, 18.5, 25.5, and 34 degrees for the 0.31-deg. sample and around 30, 31, 36 and 42.5 degrees for the 1.09-deg. sample. Least affected by the variation of the incidence angle are the diffraction spectra for the 1.09-deg sample. Changes are somewhat stronger for the 0.31-deg sample diffraction spectra, for which the detection angles $\alpha'$ were smaller. The peak maxima shift towards higher energies and the background on the high-energy side of the spectrum rises, while the overall shape of the spectrum is conserved. Much more pronounced changes of the spectral shape are found for the reflectivity signal and they occur in the same way for both samples. For comparison the reflectivity signal of the 0.31~ sample at 5~deg of incidence has been plotted also as the dotted line in the right diagram of Fig. \ref{data}.

\begin{figure}[t]
\begin{center}
\includegraphics[width=0.45\textwidth]{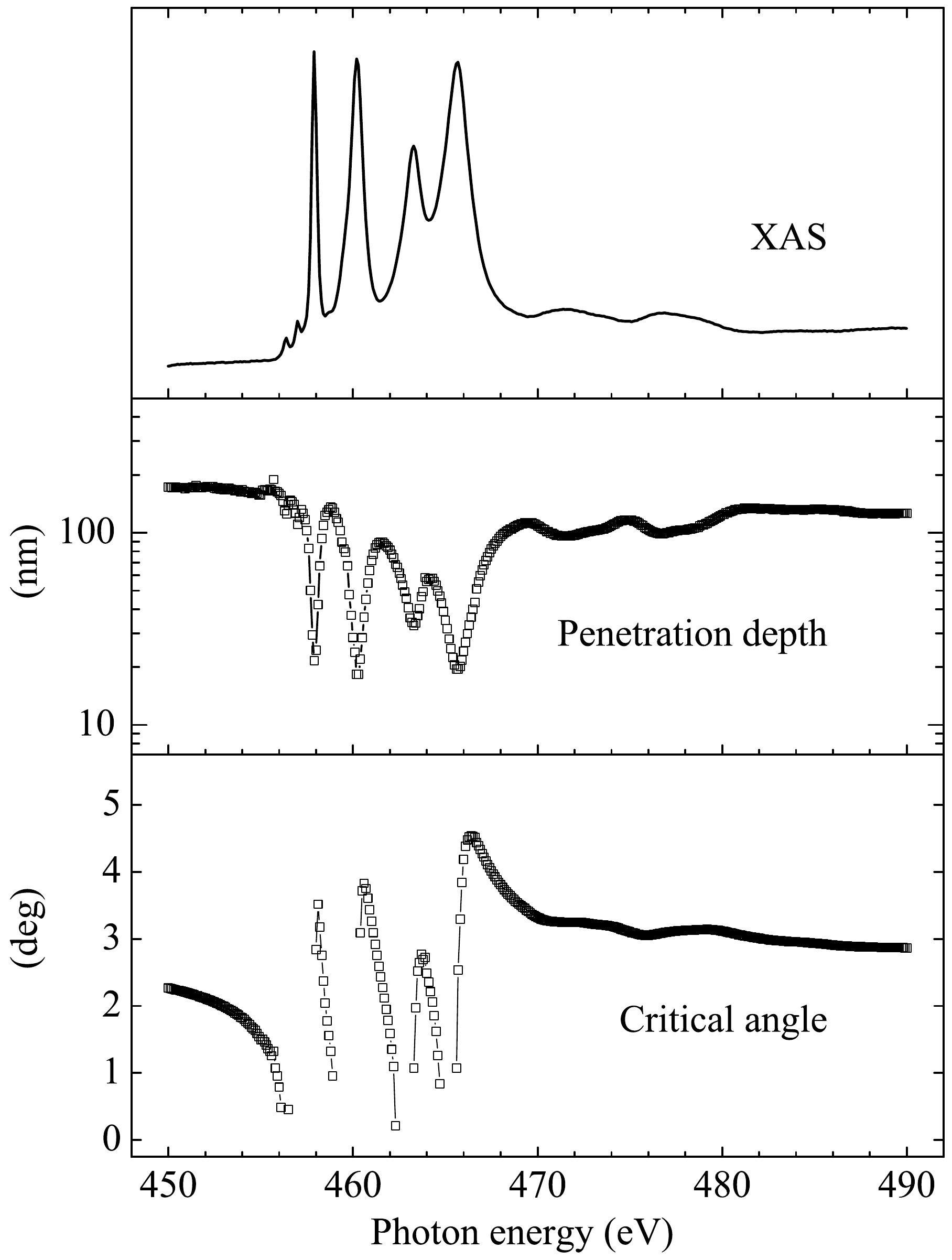}
\caption{XAS spectrum (top), penetration depth of photons $\Lambda$ on a logarithmic
scale (middle) and the critical angle $\alpha_c$ (bottom), the latter two obtained from the
optical constants presented in Fig.~\ref{DeltaBeta}.}\label{critAngLamb}
\end{center}
\end{figure}

The strong changes of $\delta$ across resonance lead to large variations of the angle of total reflection, $\alpha_c$. The changes of $\beta$ affect the photon penetration depth $\Lambda$.
Fig.~\ref{critAngLamb} shows these quantities in comparison with the XAS data across the Ti $L_{2,3}$ resonance in SrTiO$_3$. $\Lambda$ and $\alpha_c$ can be directly obtained from the optical constants; $\Lambda = 1/\mu$ is the inverse of $4 \pi \beta / \lambda$ and $\alpha_c$ is given by $\cos\alpha_c = \sqrt{1 - \delta}$. Below and above the resonance energies $\Lambda$ has values well above 100~nm, meaning that the signal is clearly bulk sensitive. At the resonance maxima the photon penetration depth drops to 20~nm, which is moderately surface sensitive. Even at its minimum value $\Lambda$ is one order of magnitude larger than the penetration depth in typical surface-sensitive techniques involving low-energy electrons like photoelectron spectroscopy. 

The critical angle varies across the resonance, resembling the behavior of $\delta$. Outside resonance  $\alpha_c$ has a value between 2 and 3~degrees. Above each resonance maximum it reaches peak values, the highest being 4.5~deg for 478~eV photon energy at the high energy side of the resonance spectrum. For energies below resonance maxima, where $\delta$ becomes negative, $\alpha_c$ is not defined and the effect of total external reflection disappears completely.

\begin{figure}[t]
\begin{center}
\includegraphics[clip,width=0.45\textwidth]{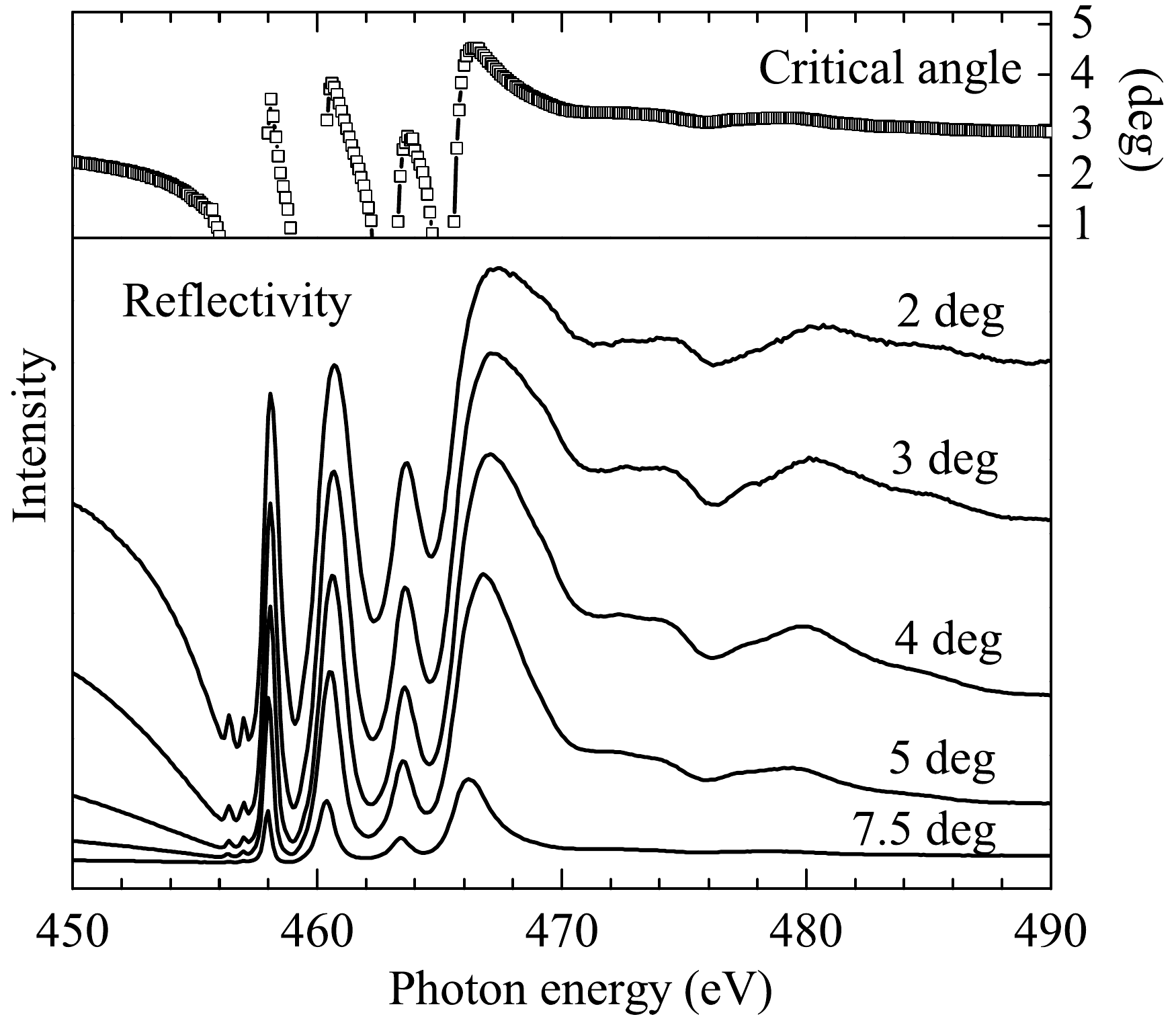}
\caption{Reflectivity for low incidence angles of the flat
SrTiO$_3$ sample (lower frame). The upper frame shows the critical angle
calculated from $\delta$.}\label{critANG}
\end{center}
\end{figure}

This rather drastic change of $\alpha_c$ with energy can be expected to strongly affect the optical spectra measured at low incidence angles. Fig.~\ref{critANG} shows specular reflectivity data for $\alpha$ between 7.5 and 2 degrees. The top diagram displays $\alpha_c$ for comparison. The reflectivity data are all plotted on the same scale. The curve measured at $\alpha=7.5$~deg resembles very much specular reflectivity for high incidence angles, but as the angle of incidence becomes smaller, the shape of the spectra changes considerably. Approaching the critical angle, the positions of the main peaks shift toward higher energies and the relative intensities of the spectrum features change; the intensity outside resonance increases strongly in comparison to the main peaks and smaller features of the spectra become more visible. Comparison with the values of the critical angle $\alpha_c$ at energies where these changes are particularly strong makes clear that these effects are caused by the occurrence of total reflection. As the incidence angle approaches $\alpha_c$, the intensity of specular reflectivity increases strongly, e.g. at the position of the highest energy peak (466~eV) for $\alpha=4.5$~deg and at the background below the resonance edge for $\alpha=2$~deg. When the incidence angle becomes even smaller than the critical angle, the increase in intensity is reduced, because total reflection has already been reached. 

\begin{figure}[t]
\begin{center}
\includegraphics[clip,width=0.45\textwidth]{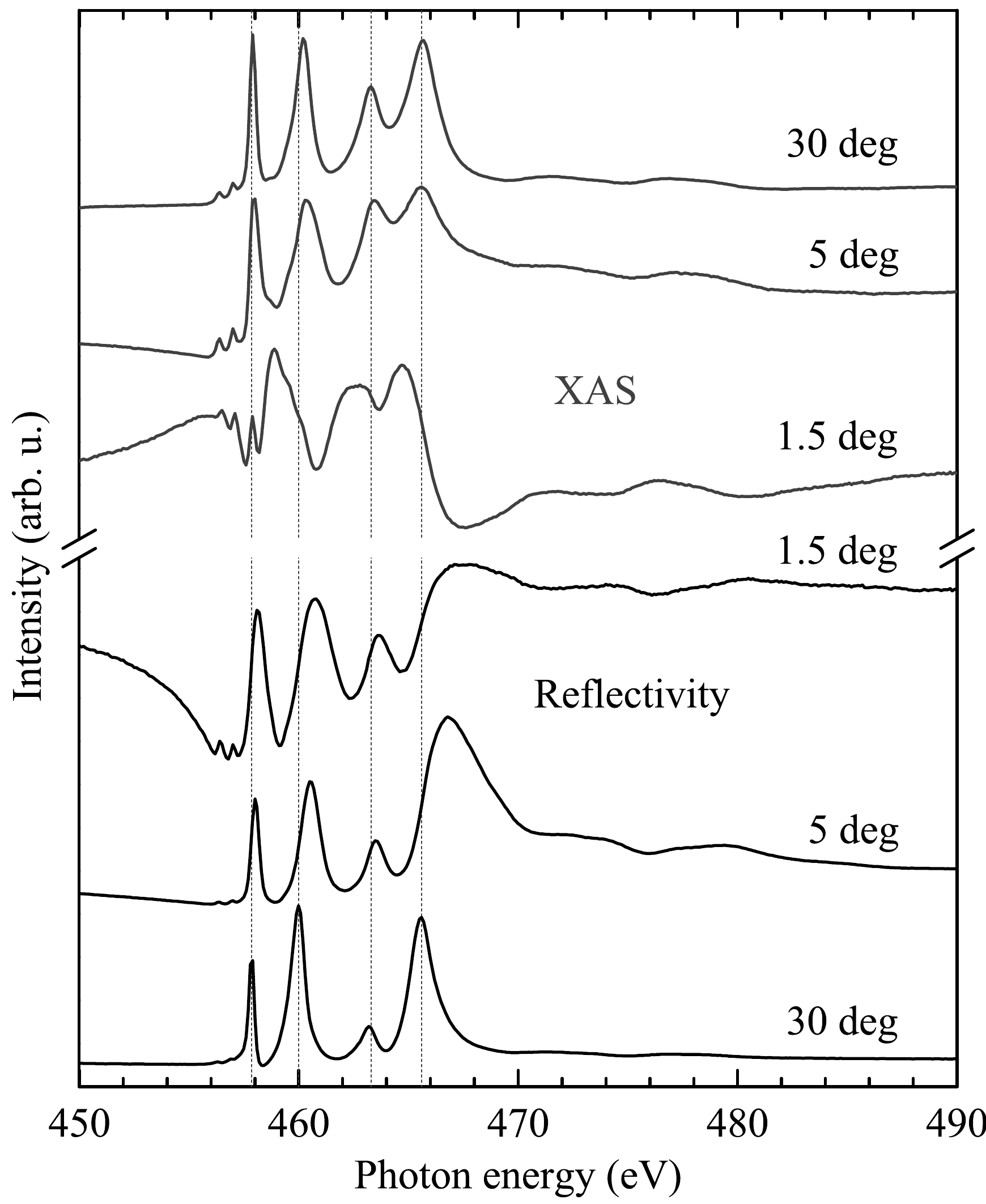}
\caption{Comparison of XAS (upper 3 curves) and reflectivity data
for different angles of incidence ($\alpha = 30°, 5°$ and
$1.5°$).}\label{ReflXAS}
\end{center}
\end{figure}

The influence of $\alpha_c$ on the absorption spectra (XAS) is even more dramatic. Since only those photons can be absorbed that can penetrate the surface, a high reflectivity suppresses the XAS signal. \cite{alders:97a,laan:88a} This effect leads to a distortion of the spectra and dominates at low incidence angles the shape of the XAS spectra, which become the negative of the reflectivity spectra.
This fact is most obvious for the data measured at $1.5$~deg incidence in the center of Fig.~\ref{ReflXAS} with minima in the absorption spectrum exactly at the positions of the maxima in the reflectivity signal. Well above the critical angle both spectra look very similar (see 30~deg of incidence). These results show that the effect of total reflection indeed dominates both, reflectivity and absorption spectra, for low incidence angles.

\section{Summary}

The stepped surface of a single crystal can work as an artificial 2-dimensional superstructure, with a periodicity in the nm-range, which can be used to produce diffraction peaks with soft x-rays. The signal is well visible, even for a step density smaller than 1/50 of the topmost layer, and can be further amplified if energies corresponding to a resonance edge of the system are used. This demonstrates again that resonant soft x-ray diffraction is extremely sensitive and suitable to study also very ``dilute'' systems. We investigated at the Ti $L_{2,3}$ resonance the first-order diffraction signal from stepped SrTiO$_3$ surfaces with a terrace width of 20 and 70~nm. The spectrum of the diffraction signal with energy resembles the spectrum for the specular reflectivity data. The x-ray absorption spectrum differs from both in the relative intensities of the spectral features and in the energy position of the peak maxima. We determined the optical parameters across resonances from specular reflectivity and found that the deviations in the spectral shape between XAS and scattering data reflect the different quantities probed in the different signals. While XAS is probing the imaginary part of the scattering amplitude only, the two scattering signals are probing the squared norm of the scattering cross section. This leads to an enhancement of the stronger over the weaker spectral features in the scattering signals. 

It is well known that changes of the optical sample properties across resonance distorts the reflectivity and XAS spectra recorded at small incidence angles. This distortion is related to the change of the angle of total reflection with photon energy, which dominates the spectral shape of both signals at very small angles. We find the spectral shape of the resonance spectra from the step edges less affected, but also in resonant diffraction spectra energy shifts of the resonance maxima can occur, when they were recorded in grazing incidence or detection geometry.

\acknowledgments
We gratefully acknowledge the expert support and excellent working conditions at BESSY and skillful technical assistance by L. Hamdan. The research in K\"oln was supported by the Deutsche Forschungsgemeinschaft through SFB 608. G. R. and D. B. acknowledge support from the Netherlands Organization for Scientific Research (NWO).


\begin{thebibliography}{35}
\expandafter\ifx\csname natexlab\endcsname\relax\def\natexlab#1{#1}\fi
\expandafter\ifx\csname bibnamefont\endcsname\relax
  \def\bibnamefont#1{#1}\fi
\expandafter\ifx\csname bibfnamefont\endcsname\relax
  \def\bibfnamefont#1{#1}\fi
\expandafter\ifx\csname citenamefont\endcsname\relax
  \def\citenamefont#1{#1}\fi
\expandafter\ifx\csname url\endcsname\relax
  \def\url#1{\texttt{#1}}\fi
\expandafter\ifx\csname urlprefix\endcsname\relax\def\urlprefix{URL }\fi
\providecommand{\bibinfo}[2]{#2}
\providecommand{\eprint}[2][]{\url{#2}}

\bibitem[{\citenamefont{Imada et~al.}(1998)\citenamefont{Imada, Fujimori, and
  Tokura}}]{Imada1998}
\bibinfo{author}{\bibfnamefont{M.}~\bibnamefont{Imada}},
  \bibinfo{author}{\bibfnamefont{A.}~\bibnamefont{Fujimori}}, \bibnamefont{and}
  \bibinfo{author}{\bibfnamefont{Y.}~\bibnamefont{Tokura}},
  \bibinfo{journal}{Rev. Mod. Phys.} \textbf{\bibinfo{volume}{70}},
  \bibinfo{pages}{1039} (\bibinfo{year}{1998}).

\bibitem[{\citenamefont{Wilkins
  et~al.}(2003{\natexlab{a}})\citenamefont{Wilkins, Hatton, Roper, Prabhakaran,
  and Boothroyd}}]{Wilkins2003_1}
\bibinfo{author}{\bibfnamefont{S.~B.} \bibnamefont{Wilkins}},
  \bibinfo{author}{\bibfnamefont{P.~D.} \bibnamefont{Hatton}},
  \bibinfo{author}{\bibfnamefont{M.~D.} \bibnamefont{Roper}},
  \bibinfo{author}{\bibfnamefont{D.}~\bibnamefont{Prabhakaran}},
  \bibnamefont{and} \bibinfo{author}{\bibfnamefont{A.~T.}
  \bibnamefont{Boothroyd}}, \bibinfo{journal}{Phys. Rev. Lett.}
  \textbf{\bibinfo{volume}{90}}, \bibinfo{eid}{187201}
  (\bibinfo{year}{2003}{\natexlab{a}}).

\bibitem[{\citenamefont{Wilkins
  et~al.}(2003{\natexlab{b}})\citenamefont{Wilkins, Spencer, Hatton, Collins,
  Roper, Prabhakaran, and Boothroyd}}]{Wilkins2003_2}
\bibinfo{author}{\bibfnamefont{S.~B.} \bibnamefont{Wilkins}},
  \bibinfo{author}{\bibfnamefont{P.~D.} \bibnamefont{Spencer}},
  \bibinfo{author}{\bibfnamefont{P.~D.} \bibnamefont{Hatton}},
  \bibinfo{author}{\bibfnamefont{S.~P.} \bibnamefont{Collins}},
  \bibinfo{author}{\bibfnamefont{M.~D.} \bibnamefont{Roper}},
  \bibinfo{author}{\bibfnamefont{D.}~\bibnamefont{Prabhakaran}},
  \bibnamefont{and} \bibinfo{author}{\bibfnamefont{A.~T.}
  \bibnamefont{Boothroyd}}, \bibinfo{journal}{Phys. Rev. Lett.}
  \textbf{\bibinfo{volume}{91}}, \bibinfo{pages}{167205}
  (\bibinfo{year}{2003}{\natexlab{b}}).

\bibitem[{\citenamefont{Dhesi et~al.}(2004)\citenamefont{Dhesi, Mirone, DeNadai,
  Ohresser, Bencok, Brookes, Reutler, Revcolevschi, Tagliaferri, Toulemonde, 
  van der Laan}}]{Dhesi2004}
\bibinfo{author}{\bibfnamefont{S.~S.} \bibnamefont{Dhesi}},
  \bibinfo{author}{\bibfnamefont{A.}~\bibnamefont{Mirone}},
  \bibinfo{author}{\bibfnamefont{C.} \bibnamefont{De Nadai}},
  \bibinfo{author}{\bibfnamefont{P.}~\bibnamefont{Ohresser}},
  \bibinfo{author}{\bibfnamefont{P.}~\bibnamefont{Bencok}},
  \bibinfo{author}{\bibfnamefont{N.~B.} \bibnamefont{Brookes}},
  \bibinfo{author}{\bibfnamefont{P.}~\bibnamefont{Reutler}},
  \bibinfo{author}{\bibfnamefont{A.}~\bibnamefont{Revcolevschi}},
  \bibinfo{author}{\bibfnamefont{A.}~\bibnamefont{Tagliaferri}},
  \bibinfo{author}{\bibfnamefont{O.}~\bibnamefont{Toulemonde}},
  \bibnamefont{and} \bibinfo{author}{\bibfnamefont{G.}
  \bibnamefont{van der Laan}}, \bibinfo{journal}{Phys. Rev. Lett.}
  \textbf{\bibinfo{volume}{92}}, \bibinfo{eid}{056403} (\bibinfo{year}{2004}).

\bibitem[{\citenamefont{Thomas et~al.}(2004)\citenamefont{Thomas, Hill,
  Grenier, Kim, Abbamonte, Venema, Rusydi, Tomioka, Tokura, McMorrow, Sawatzky, and van Veenendaal}}]{Thomas2004}
\bibinfo{author}{\bibfnamefont{K.~J.} \bibnamefont{Thomas}},
  \bibinfo{author}{\bibfnamefont{J.~P.} \bibnamefont{Hill}},
  \bibinfo{author}{\bibfnamefont{S.}~\bibnamefont{Grenier}},
  \bibinfo{author}{\bibfnamefont{Y.-J.} \bibnamefont{Kim}},
  \bibinfo{author}{\bibfnamefont{P.}~\bibnamefont{Abbamonte}},
  \bibinfo{author}{\bibfnamefont{L.}~\bibnamefont{Venema}},
  \bibinfo{author}{\bibfnamefont{A.}~\bibnamefont{Rusydi}},
  \bibinfo{author}{\bibfnamefont{Y.}~\bibnamefont{Tomioka}},
  \bibinfo{author}{\bibfnamefont{Y.}~\bibnamefont{Tokura}},
  \bibinfo{author}{\bibfnamefont{D.~F.} \bibnamefont{McMorrow}},
  \bibinfo{author}{\bibfnamefont{G.}~\bibnamefont{Sawatzky}},
  \bibnamefont{and} \bibinfo{author}{\bibfnamefont{M.}
  \bibnamefont{van Veenendaal}}, \bibinfo{journal}{Phys. Rev. Lett.}
  \textbf{\bibinfo{volume}{92}}, \bibinfo{eid}{237204} (\bibinfo{year}{2004}).

\bibitem[{\citenamefont{Wilkins et~al.}(2005)\citenamefont{Wilkins, Stojic,
  Beale, Binggeli, Castleton, Bencok, Prabhakaran, Boothroyd, Hatton, and
  Altarelli}}]{Wilkins2005}
\bibinfo{author}{\bibfnamefont{S.~B.} \bibnamefont{Wilkins}},
  \bibinfo{author}{\bibfnamefont{N.}~\bibnamefont{Stojic}},
  \bibinfo{author}{\bibfnamefont{T.~A.~W.} \bibnamefont{Beale}},
  \bibinfo{author}{\bibfnamefont{N.}~\bibnamefont{Binggeli}},
  \bibinfo{author}{\bibfnamefont{C.~W.~M.} \bibnamefont{Castleton}},
  \bibinfo{author}{\bibfnamefont{P.}~\bibnamefont{Bencok}},
  \bibinfo{author}{\bibfnamefont{D.}~\bibnamefont{Prabhakaran}},
  \bibinfo{author}{\bibfnamefont{A.~T.} \bibnamefont{Boothroyd}},
  \bibinfo{author}{\bibfnamefont{P.~D.} \bibnamefont{Hatton}},
  \bibnamefont{and}
  \bibinfo{author}{\bibfnamefont{M.}~\bibnamefont{Altarelli}},
  \bibinfo{journal}{Phys. Rev. B} \textbf{\bibinfo{volume}{71}},
  \bibinfo{eid}{245102} (\bibinfo{year}{2005}).

\bibitem[{\citenamefont{Staub et~al.}(2005)\citenamefont{Staub, Scagnoli,
  Mulders, Katsumata, Honda, Grimmer, Horisberger, and Tonnerre}}]{Staub2005}
\bibinfo{author}{\bibfnamefont{U.}~\bibnamefont{Staub}},
  \bibinfo{author}{\bibfnamefont{V.}~\bibnamefont{Scagnoli}},
  \bibinfo{author}{\bibfnamefont{A.~M.} \bibnamefont{Mulders}},
  \bibinfo{author}{\bibfnamefont{K.}~\bibnamefont{Katsumata}},
  \bibinfo{author}{\bibfnamefont{Z.}~\bibnamefont{Honda}},
  \bibinfo{author}{\bibfnamefont{H.}~\bibnamefont{Grimmer}},
  \bibinfo{author}{\bibfnamefont{M.}~\bibnamefont{Horisberger}},
  \bibnamefont{and} \bibinfo{author}{\bibfnamefont{J.~M.}
  \bibnamefont{Tonnerre}}, \bibinfo{journal}{Phys. Rev. B}
  \textbf{\bibinfo{volume}{71}}, \bibinfo{eid}{214421} (\bibinfo{year}{2005}).

\bibitem[{\citenamefont{\c{S}erban Smadici
  et~al.}(2007{\natexlab{a}})\citenamefont{\c{S}erban Smadici, Abbamonte,
  Bhattacharya, Zhai, Jiang, Rusydi, Eckstein, Bader, and Zuo}}]{smadici:07b}
\bibinfo{author}{\bibnamefont{\c{S}erban Smadici}},
  \bibinfo{author}{\bibfnamefont{P.}~\bibnamefont{Abbamonte}},
  \bibinfo{author}{\bibfnamefont{A.}~\bibnamefont{Bhattacharya}},
  \bibinfo{author}{\bibfnamefont{X.}~\bibnamefont{Zhai}},
  \bibinfo{author}{\bibfnamefont{B.}~\bibnamefont{Jiang}},
  \bibinfo{author}{\bibfnamefont{A.}~\bibnamefont{Rusydi}},
  \bibinfo{author}{\bibfnamefont{J.~N.} \bibnamefont{Eckstein}},
  \bibinfo{author}{\bibfnamefont{S.~D.} \bibnamefont{Bader}}, \bibnamefont{and}
  \bibinfo{author}{\bibfnamefont{J.-M.} \bibnamefont{Zuo}},
  \bibinfo{journal}{Phys. Rev. Lett.} \textbf{\bibinfo{volume}{99}},
  \bibinfo{pages}{196404} (\bibinfo{year}{2007}{\natexlab{a}}).

\bibitem[{\citenamefont{Wilkins et~al.}(2006)\citenamefont{Wilkins, Stojic,
  Beale, Binggeli, Hatton, Bencok, Stanescu, Mitchell, Abbamonte, and
  Altarelli}}]{wilkins:06a}
\bibinfo{author}{\bibfnamefont{S.~B.} \bibnamefont{Wilkins}},
  \bibinfo{author}{\bibfnamefont{N.}~\bibnamefont{Stojic}},
  \bibinfo{author}{\bibfnamefont{T.~A.~W.} \bibnamefont{Beale}},
  \bibinfo{author}{\bibfnamefont{N.}~\bibnamefont{Binggeli}},
  \bibinfo{author}{\bibfnamefont{P.~D.} \bibnamefont{Hatton}},
  \bibinfo{author}{\bibfnamefont{P.}~\bibnamefont{Bencok}},
  \bibinfo{author}{\bibfnamefont{S.}~\bibnamefont{Stanescu}},
  \bibinfo{author}{\bibfnamefont{J.~F.} \bibnamefont{Mitchell}},
  \bibinfo{author}{\bibfnamefont{P.}~\bibnamefont{Abbamonte}},
  \bibnamefont{and}
  \bibinfo{author}{\bibfnamefont{M.}~\bibnamefont{Altarelli}},
  \bibinfo{journal}{J. Phys. Cond Mat.} \textbf{\bibinfo{volume}{18}},
  \bibinfo{pages}{L323} (\bibinfo{year}{2006}).

\bibitem[{\citenamefont{Staub et~al.}(2006)\citenamefont{Staub, Scagnoli,
  Mulders, Janousch, Honda, and Tonnerre}}]{staub:06a}
\bibinfo{author}{\bibfnamefont{U.}~\bibnamefont{Staub}},
  \bibinfo{author}{\bibfnamefont{V.}~\bibnamefont{Scagnoli}},
  \bibinfo{author}{\bibfnamefont{A.~M.} \bibnamefont{Mulders}},
  \bibinfo{author}{\bibfnamefont{M.}~\bibnamefont{Janousch}},
  \bibinfo{author}{\bibfnamefont{Z.}~\bibnamefont{Honda}}, \bibnamefont{and}
  \bibinfo{author}{\bibfnamefont{J.~M.} \bibnamefont{Tonnerre}},
  \bibinfo{journal}{Europhys. Lett.} \textbf{\bibinfo{volume}{76}},
  \bibinfo{pages}{926} (\bibinfo{year}{2006}).

\bibitem[{\citenamefont{Grenier
  et~al.}(2007{\natexlab{a}})\citenamefont{Grenier, Kiryukhin, Cheong, Kim,
  Hill, Thomas, Tonnerre, Joly, Staub, and Scagnoli}}]{grenier:07a}
\bibinfo{author}{\bibfnamefont{S.}~\bibnamefont{Grenier}},
  \bibinfo{author}{\bibfnamefont{V.}~\bibnamefont{Kiryukhin}},
  \bibinfo{author}{\bibfnamefont{S.-W.} \bibnamefont{Cheong}},
  \bibinfo{author}{\bibfnamefont{B.~G.} \bibnamefont{Kim}},
  \bibinfo{author}{\bibfnamefont{J.~P.} \bibnamefont{Hill}},
  \bibinfo{author}{\bibfnamefont{K.~J.} \bibnamefont{Thomas}},
  \bibinfo{author}{\bibfnamefont{J.~M.} \bibnamefont{Tonnerre}},
  \bibinfo{author}{\bibfnamefont{Y.}~\bibnamefont{Joly}},
  \bibinfo{author}{\bibfnamefont{U.}~\bibnamefont{Staub}}, \bibnamefont{and}
  \bibinfo{author}{\bibfnamefont{V.}~\bibnamefont{Scagnoli}},
  \bibinfo{journal}{Phys. Rev. B} \textbf{\bibinfo{volume}{75}},
  \bibinfo{pages}{085101} (\bibinfo{year}{2007}{\natexlab{a}}).

\bibitem[{\citenamefont{Grenier
  et~al.}(2007{\natexlab{b}})\citenamefont{Grenier, Thomas, Hill, Staub,
  Bodenthin, Garc\'{\i}a-Fern\'{a}ndez, Scagnoli, Kiryukhin, Cheong, Kim,
 and Tonnerre}}]{grenier:07b}
\bibinfo{author}{\bibfnamefont{S.}~\bibnamefont{Grenier}},
  \bibinfo{author}{\bibfnamefont{K.~J.} \bibnamefont{Thomas}},
  \bibinfo{author}{\bibfnamefont{J.~P.} \bibnamefont{Hill}},
  \bibinfo{author}{\bibfnamefont{U.}~\bibnamefont{Staub}},
  \bibinfo{author}{\bibfnamefont{Y.}~\bibnamefont{Bodenthin}},
  \bibinfo{author}{\bibfnamefont{M.}~\bibnamefont{Garc\'{\i}a-Fern\'{a}ndez}},
  \bibinfo{author}{\bibfnamefont{V.}~\bibnamefont{Scagnoli}},
  \bibinfo{author}{\bibfnamefont{V.}~\bibnamefont{Kiryukhin}},
  \bibinfo{author}{\bibfnamefont{S.-W.} \bibnamefont{Cheong}},
  \bibinfo{author}{\bibfnamefont{B.~G.} \bibnamefont{Kim}}, \bibnamefont{and}
  \bibinfo{author}{\bibfnamefont{J.~M.}~\bibnamefont{Tonnerre}},
  \bibinfo{journal}{Phys. Rev. Lett.}
  \textbf{\bibinfo{volume}{99}}, \bibinfo{pages}{206403}
  (\bibinfo{year}{2007}{\natexlab{b}}).

\bibitem[{\citenamefont{Bodenthin et~al.}(2008)\citenamefont{Bodenthin, Staub,
  Garc\'{\i}a-Fern\'{a}ndez, Janoschek, Schlappa, Golovenchits, Sanina, and
  Lushnikov}}]{bodenthin:08a}
\bibinfo{author}{\bibfnamefont{Y.}~\bibnamefont{Bodenthin}},
  \bibinfo{author}{\bibfnamefont{U.}~\bibnamefont{Staub}},
  \bibinfo{author}{\bibfnamefont{M.}~\bibnamefont{Garc\'{\i}a-Fern\'{a}ndez}},
  \bibinfo{author}{\bibfnamefont{M.}~\bibnamefont{Janoschek}},
  \bibinfo{author}{\bibfnamefont{J.}~\bibnamefont{Schlappa}},
  \bibinfo{author}{\bibfnamefont{E.~I.} \bibnamefont{Golovenchits}},
  \bibinfo{author}{\bibfnamefont{V.~A.} \bibnamefont{Sanina}},
  \bibnamefont{and} \bibinfo{author}{\bibfnamefont{S.~G.}
  \bibnamefont{Lushnikov}}, \bibinfo{journal}{Phys. Rev. Lett.}
  \textbf{\bibinfo{volume}{100}}, \bibinfo{pages}{027201}
  (\bibinfo{year}{2008}).

\bibitem[{\citenamefont{Garc\'{\i}a-Fern\'{a}ndez
  et~al.}(2008)\citenamefont{Garc\'{\i}a-Fern\'{a}ndez, Staub, Bodenthin,
  Lawrence, Mulders, Buckley, Weyeneth, Pomjakushina, and Conder}}]{garcia:08a}
\bibinfo{author}{\bibfnamefont{M.}~\bibnamefont{Garc\'{\i}a-Fern\'{a}ndez}},
  \bibinfo{author}{\bibfnamefont{U.}~\bibnamefont{Staub}},
  \bibinfo{author}{\bibfnamefont{Y.}~\bibnamefont{Bodenthin}},
  \bibinfo{author}{\bibfnamefont{S.~M.} \bibnamefont{Lawrence}},
  \bibinfo{author}{\bibfnamefont{A.~M.} \bibnamefont{Mulders}},
  \bibinfo{author}{\bibfnamefont{C.~E.} \bibnamefont{Buckley}},
  \bibinfo{author}{\bibfnamefont{S.}~\bibnamefont{Weyeneth}},
  \bibinfo{author}{\bibfnamefont{E.}~\bibnamefont{Pomjakushina}},
  \bibnamefont{and} \bibinfo{author}{\bibfnamefont{K.}~\bibnamefont{Conder}},
  \bibinfo{journal}{Phys. Rev. B} \textbf{\bibinfo{volume}{77}},
  \bibinfo{pages}{060402(R)} (\bibinfo{year}{2008}).

\bibitem[{\citenamefont{Huang et~al.}(2006)\citenamefont{Huang, Lin, Okamoto,
  Chao, Jeng, Guo, Hsu, Huang, Ling, Wu, Yang, and Chen}}]{Huang2006}
\bibinfo{author}{\bibfnamefont{D.~J.} \bibnamefont{Huang}},
  \bibinfo{author}{\bibfnamefont{H.-J.} \bibnamefont{Lin}},
  \bibinfo{author}{\bibfnamefont{J.}~\bibnamefont{Okamoto}},
  \bibinfo{author}{\bibfnamefont{K.~S.} \bibnamefont{Chao}},
  \bibinfo{author}{\bibfnamefont{H.-T.} \bibnamefont{Jeng}},
  \bibinfo{author}{\bibfnamefont{G.~Y.} \bibnamefont{Guo}},
  \bibinfo{author}{\bibfnamefont{C.-H.} \bibnamefont{Hsu}},
  \bibinfo{author}{\bibfnamefont{C.-M.} \bibnamefont{Huang}},
  \bibinfo{author}{\bibfnamefont{D.~C.} \bibnamefont{Ling}},
  \bibinfo{author}{\bibfnamefont{W.~B.} \bibnamefont{Wu}},
  \bibinfo{author}{\bibfnamefont{C.~S.}~\bibnamefont{Yang}},
  \bibnamefont{and} \bibinfo{author}{\bibfnamefont{C.~T.}~\bibnamefont{Chen}},    
  \bibinfo{journal}{Phys. Rev. Lett.}
  \textbf{\bibinfo{volume}{96}}, \bibinfo{eid}{096401} (\bibinfo{year}{2006}).

\bibitem[{\citenamefont{Schlappa et~al.}(2008)\citenamefont{Schlappa,
  Sch{\"u}{\ss}ler-Langeheine, Chang, Ott, Tanaka, Hu, Haverkort, Schierle,
  Weschke, Kaindl, and Tjeng}}]{schlappa:08a}
\bibinfo{author}{\bibfnamefont{J.}~\bibnamefont{Schlappa}},
  \bibinfo{author}{\bibfnamefont{C.}~\bibnamefont{Sch{\"u}{\ss}ler-Langeheine}%
}, \bibinfo{author}{\bibfnamefont{C.~F.} \bibnamefont{Chang}},
  \bibinfo{author}{\bibfnamefont{H.}~\bibnamefont{Ott}},
  \bibinfo{author}{\bibfnamefont{A.}~\bibnamefont{Tanaka}},
  \bibinfo{author}{\bibfnamefont{Z.}~\bibnamefont{Hu}},
  \bibinfo{author}{\bibfnamefont{M.~W.} \bibnamefont{Haverkort}},
  \bibinfo{author}{\bibfnamefont{E.}~\bibnamefont{Schierle}},
  \bibinfo{author}{\bibfnamefont{E.}~\bibnamefont{Weschke}},
  \bibinfo{author}{\bibfnamefont{G.}~\bibnamefont{Kaindl}},
  \bibnamefont{and} \bibinfo{author}{\bibfnamefont{L.~H.}~\bibnamefont{Tjeng}},
  \bibinfo{journal}{Phys. Rev. Lett.}
  \textbf{\bibinfo{volume}{100}}, \bibinfo{pages}{026406}
  (\bibinfo{year}{2008}).

\bibitem[{\citenamefont{Sch\"u{\ss}ler-Langeheine
  et~al.}(2005)\citenamefont{Sch\"u{\ss}ler-Langeheine, Schlappa, Tanaka, Hu,
  Chang, Schierle, Benomar, Ott, Weschke, Kaindl, Friedt, Sawatzky, Lin, Chen, Braden, and Tjeng}}]{ourPRL2005}
\bibinfo{author}{\bibfnamefont{C.}~\bibnamefont{Sch\"u{\ss}ler-Langeheine}},
  \bibinfo{author}{\bibfnamefont{J.}~\bibnamefont{Schlappa}},
  \bibinfo{author}{\bibfnamefont{A.}~\bibnamefont{Tanaka}},
  \bibinfo{author}{\bibfnamefont{Z.}~\bibnamefont{Hu}},
  \bibinfo{author}{\bibfnamefont{C.~F.} \bibnamefont{Chang}},
  \bibinfo{author}{\bibfnamefont{E.}~\bibnamefont{Schierle}},
  \bibinfo{author}{\bibfnamefont{M.}~\bibnamefont{Benomar}},
  \bibinfo{author}{\bibfnamefont{H.}~\bibnamefont{Ott}},
  \bibinfo{author}{\bibfnamefont{E.}~\bibnamefont{Weschke}},
  \bibinfo{author}{\bibfnamefont{G.}~\bibnamefont{Kaindl}},
  \bibinfo{author}{\bibfnamefont{O.}~\bibnamefont{Friedt}},
  \bibinfo{author}{\bibfnamefont{G.~A.}~\bibnamefont{Sawatzky}},
  \bibinfo{author}{\bibfnamefont{H.-J.}~\bibnamefont{Lin}},
  \bibinfo{author}{\bibfnamefont{C.-T.}~\bibnamefont{Chen}},
  \bibinfo{author}{\bibfnamefont{M.} \bibnamefont{Braden}},
  \bibnamefont{and} \bibinfo{author}{\bibfnamefont{L.~H.}~\bibnamefont{Tjeng}},
  \bibinfo{journal}{Phys. Rev. Lett.}
  \textbf{\bibinfo{volume}{95}}, \bibinfo{pages}{156402}
  (\bibinfo{year}{2005}).

\bibitem[{\citenamefont{Scagnoli et~al.}(2006)\citenamefont{Scagnoli, Staub,
  Mulders, Janousch, Meijer, Hammerl, Tonnerre, and Stojic}}]{Scagnoli2006}
\bibinfo{author}{\bibfnamefont{V.}~\bibnamefont{Scagnoli}},
  \bibinfo{author}{\bibfnamefont{U.}~\bibnamefont{Staub}},
  \bibinfo{author}{\bibfnamefont{A.~M.} \bibnamefont{Mulders}},
  \bibinfo{author}{\bibfnamefont{M.}~\bibnamefont{Janousch}},
  \bibinfo{author}{\bibfnamefont{G.~I.} \bibnamefont{Meijer}},
  \bibinfo{author}{\bibfnamefont{G.}~\bibnamefont{Hammerl}},
  \bibinfo{author}{\bibfnamefont{J.~M.} \bibnamefont{Tonnerre}},
  \bibnamefont{and} \bibinfo{author}{\bibfnamefont{N.}~\bibnamefont{Stojic}},
  \bibinfo{journal}{Phys. Rev. B} \textbf{\bibinfo{volume}{73}},
  \bibinfo{eid}{100409(R)} (\bibinfo{year}{2006}).

\bibitem[{\citenamefont{Staub et~al.}(2007)\citenamefont{Staub,
  Garc\'{\i}a-Fern\'{a}ndez, Mulders, Bodenthin, Mart\'inez-Lope, and
  Alonso}}]{staub:07a}
\bibinfo{author}{\bibfnamefont{U.}~\bibnamefont{Staub}},
  \bibinfo{author}{\bibfnamefont{M.}~\bibnamefont{Garc\'{\i}a-Fern\'{a}ndez}},
  \bibinfo{author}{\bibfnamefont{A.~M.} \bibnamefont{Mulders}},
  \bibinfo{author}{\bibfnamefont{Y.}~\bibnamefont{Bodenthin}},
  \bibinfo{author}{\bibfnamefont{M.~J.} \bibnamefont{Mart\'inez-Lope}},
  \bibnamefont{and} \bibinfo{author}{\bibfnamefont{J.~A.}
  \bibnamefont{Alonso}}, \bibinfo{journal}{J. Phys. Cond Mat.}
  \textbf{\bibinfo{volume}{19}}, \bibinfo{pages}{092201}
  (\bibinfo{year}{2007}).

\bibitem[{\citenamefont{Abbamonte et~al.}(2002)\citenamefont{Abbamonte, Venema,
  Rusydi, Sawatzky, Logvenov, and Bozovic}}]{Abbamonte2002}
\bibinfo{author}{\bibfnamefont{P.}~\bibnamefont{Abbamonte}},
  \bibinfo{author}{\bibfnamefont{L.}~\bibnamefont{Venema}},
  \bibinfo{author}{\bibfnamefont{A.}~\bibnamefont{Rusydi}},
  \bibinfo{author}{\bibfnamefont{G.~A.} \bibnamefont{Sawatzky}},
  \bibinfo{author}{\bibfnamefont{G.}~\bibnamefont{Logvenov}}, \bibnamefont{and}
  \bibinfo{author}{\bibfnamefont{I.}~\bibnamefont{Bozovic}},
  \bibinfo{journal}{Science} \textbf{\bibinfo{volume}{297}},
  \bibinfo{pages}{581} (\bibinfo{year}{2002}).

\bibitem[{\citenamefont{Abbamonte et~al.}(2004)\citenamefont{Abbamonte,
  Blumberg, Rusydi, Gozar, Evans, Siegrist, Venema, Eisaki, Isaacs, and
  Sawatzky}}]{Abbamonte2004}
\bibinfo{author}{\bibfnamefont{P.}~\bibnamefont{Abbamonte}},
  \bibinfo{author}{\bibfnamefont{G.}~\bibnamefont{Blumberg}},
  \bibinfo{author}{\bibfnamefont{A.}~\bibnamefont{Rusydi}},
  \bibinfo{author}{\bibfnamefont{A.}~\bibnamefont{Gozar}},
  \bibinfo{author}{\bibfnamefont{P.~G.} \bibnamefont{Evans}},
  \bibinfo{author}{\bibfnamefont{T.}~\bibnamefont{Siegrist}},
  \bibinfo{author}{\bibfnamefont{L.}~\bibnamefont{Venema}},
  \bibinfo{author}{\bibfnamefont{H.}~\bibnamefont{Eisaki}},
  \bibinfo{author}{\bibfnamefont{E.~D.} \bibnamefont{Isaacs}},
  \bibnamefont{and} \bibinfo{author}{\bibfnamefont{G.~A.}
  \bibnamefont{Sawatzky}}, \bibinfo{journal}{Nature}
  \textbf{\bibinfo{volume}{431}}, \bibinfo{pages}{1078} (\bibinfo{year}{2004}).

\bibitem[{\citenamefont{Abbamonte et~al.}(2005)\citenamefont{Abbamonte, Rusydi,
  Smadici, Gu, Sawatzky, and Feng}}]{Abbamonte2005}
\bibinfo{author}{\bibfnamefont{P.}~\bibnamefont{Abbamonte}},
  \bibinfo{author}{\bibfnamefont{A.}~\bibnamefont{Rusydi}},
  \bibinfo{author}{\bibfnamefont{S.}~\bibnamefont{Smadici}},
  \bibinfo{author}{\bibfnamefont{G.~D.} \bibnamefont{Gu}},
  \bibinfo{author}{\bibfnamefont{G.~A.} \bibnamefont{Sawatzky}},
  \bibnamefont{and} \bibinfo{author}{\bibfnamefont{D.~L.} \bibnamefont{Feng}},
  \bibinfo{journal}{Nature Physics} \textbf{\bibinfo{volume}{1}},
  \bibinfo{pages}{155} (\bibinfo{year}{2005}).

\bibitem[{\citenamefont{Rusydi et~al.}(2006)\citenamefont{Rusydi, Abbamonte,
  Eisaki, Fujimaki, Blumberg, Uchida, and Sawatzky}}]{Rusydi2006}
\bibinfo{author}{\bibfnamefont{A.}~\bibnamefont{Rusydi}},
  \bibinfo{author}{\bibfnamefont{P.}~\bibnamefont{Abbamonte}},
  \bibinfo{author}{\bibfnamefont{H.}~\bibnamefont{Eisaki}},
  \bibinfo{author}{\bibfnamefont{Y.}~\bibnamefont{Fujimaki}},
  \bibinfo{author}{\bibfnamefont{G.}~\bibnamefont{Blumberg}},
  \bibinfo{author}{\bibfnamefont{S.}~\bibnamefont{Uchida}}, \bibnamefont{and}
  \bibinfo{author}{\bibfnamefont{G.~A.} \bibnamefont{Sawatzky}},
  \bibinfo{journal}{Phys. Rev. Lett.} \textbf{\bibinfo{volume}{97}},
  \bibinfo{eid}{016403} (\bibinfo{year}{2006}).

\bibitem[{\citenamefont{Rusydi et~al.}(2007)\citenamefont{Rusydi, Berciu,
  Abbamonte, Smadici, Eisaki, Fujimaki, Uchida, R\"{u}bhausen, and
  Sawatzky}}]{rusydi:07a}
\bibinfo{author}{\bibfnamefont{A.}~\bibnamefont{Rusydi}},
  \bibinfo{author}{\bibfnamefont{M.}~\bibnamefont{Berciu}},
  \bibinfo{author}{\bibfnamefont{P.}~\bibnamefont{Abbamonte}},
  \bibinfo{author}{\bibfnamefont{S.}~\bibnamefont{Smadici}},
  \bibinfo{author}{\bibfnamefont{H.}~\bibnamefont{Eisaki}},
  \bibinfo{author}{\bibfnamefont{Y.}~\bibnamefont{Fujimaki}},
  \bibinfo{author}{\bibfnamefont{S.}~\bibnamefont{Uchida}},
  \bibinfo{author}{\bibfnamefont{M.}~\bibnamefont{R\"{u}bhausen}},
  \bibnamefont{and} \bibinfo{author}{\bibfnamefont{G.~A.}
  \bibnamefont{Sawatzky}}, \bibinfo{journal}{Phys. Rev. B}
  \textbf{\bibinfo{volume}{75}}, \bibinfo{pages}{104510}
  (\bibinfo{year}{2007}).

\bibitem[{\citenamefont{\c{S}erban Smadici
  et~al.}(2007{\natexlab{b}})\citenamefont{\c{S}erban Smadici, Abbamonte,
  Taguchi, Kohsaka, Sasagawa, Azuma, Takano, and Takagi}}]{smadici:07a}
\bibinfo{author}{\bibnamefont{\c{S}erban Smadici}},
  \bibinfo{author}{\bibfnamefont{P.}~\bibnamefont{Abbamonte}},
  \bibinfo{author}{\bibfnamefont{M.}~\bibnamefont{Taguchi}},
  \bibinfo{author}{\bibfnamefont{Y.}~\bibnamefont{Kohsaka}},
  \bibinfo{author}{\bibfnamefont{T.}~\bibnamefont{Sasagawa}},
  \bibinfo{author}{\bibfnamefont{M.}~\bibnamefont{Azuma}},
  \bibinfo{author}{\bibfnamefont{M.}~\bibnamefont{Takano}}, \bibnamefont{and}
  \bibinfo{author}{\bibfnamefont{H.}~\bibnamefont{Takagi}},
  \bibinfo{journal}{Phys. Rev. B} \textbf{\bibinfo{volume}{75}},
  \bibinfo{pages}{075104} (\bibinfo{year}{2007}{\natexlab{b}}).

\bibitem[{\citenamefont{Rusydi et~al.}(2008)\citenamefont{Rusydi, Abbamonte,
  Eisaki, Fujimaki, Smadici, Motoyama, Uchida, Kim, R\"{u}bhausen, and
  Sawatzky}}]{rusydi:08a}
\bibinfo{author}{\bibfnamefont{A.}~\bibnamefont{Rusydi}},
  \bibinfo{author}{\bibfnamefont{P.}~\bibnamefont{Abbamonte}},
  \bibinfo{author}{\bibfnamefont{H.}~\bibnamefont{Eisaki}},
  \bibinfo{author}{\bibfnamefont{Y.}~\bibnamefont{Fujimaki}},
  \bibinfo{author}{\bibfnamefont{S.}~\bibnamefont{Smadici}},
  \bibinfo{author}{\bibfnamefont{N.}~\bibnamefont{Motoyama}},
  \bibinfo{author}{\bibfnamefont{S.}~\bibnamefont{Uchida}},
  \bibinfo{author}{\bibfnamefont{Y.-J.} \bibnamefont{Kim}},
  \bibinfo{author}{\bibfnamefont{M.}~\bibnamefont{R\"{u}bhausen}},
  \bibnamefont{and} \bibinfo{author}{\bibfnamefont{G.~A.}
  \bibnamefont{Sawatzky}}, \bibinfo{journal}{Phys. Rev. Lett.}
  \textbf{\bibinfo{volume}{100}}, \bibinfo{pages}{036403}
  (\bibinfo{year}{2008}).

\bibitem[{\citenamefont{Zegkinoglou et~al.}(2005)\citenamefont{Zegkinoglou,
  Strempfer, Nelson, Hill, Chakhalian, Bernhard, Lang, Srajer, Fukazawa,
  Nakatsuji et~al.}}]{Zegkinoglou2005}
\bibinfo{author}{\bibfnamefont{I.}~\bibnamefont{Zegkinoglou}},
  \bibinfo{author}{\bibfnamefont{J.}~\bibnamefont{Strempfer}},
  \bibinfo{author}{\bibfnamefont{C.~S.} \bibnamefont{Nelson}},
  \bibinfo{author}{\bibfnamefont{J.~P.} \bibnamefont{Hill}},
  \bibinfo{author}{\bibfnamefont{J.}~\bibnamefont{Chakhalian}},
  \bibinfo{author}{\bibfnamefont{C.}~\bibnamefont{Bernhard}},
  \bibinfo{author}{\bibfnamefont{J.~C.} \bibnamefont{Lang}},
  \bibinfo{author}{\bibfnamefont{G.}~\bibnamefont{Srajer}},
  \bibinfo{author}{\bibfnamefont{H.}~\bibnamefont{Fukazawa}},
  \bibinfo{author}{\bibfnamefont{S.}~\bibnamefont{Nakatsuji}},
  \bibinfo{author}{\bibfnamefont{Y.}~\bibnamefont{Maeno}},
  \bibnamefont{and} \bibinfo{author}{\bibfnamefont{B.}
  \bibnamefont{Keimer}}, \bibinfo{journal}{Phys. Rev. Lett.}
  \textbf{\bibinfo{volume}{95}}, \bibinfo{eid}{136401} (\bibinfo{year}{2005}).

\bibitem[{\citenamefont{de~Groot et~al.}(1990)\citenamefont{de~Groot, Fuggle,
  Thole, and Sawatzky}}]{deGroot1990}
\bibinfo{author}{\bibfnamefont{F.~M.~F.} \bibnamefont{de~Groot}},
  \bibinfo{author}{\bibfnamefont{J.~C.} \bibnamefont{Fuggle}},
  \bibinfo{author}{\bibfnamefont{B.~T.} \bibnamefont{Thole}}, \bibnamefont{and}
  \bibinfo{author}{\bibfnamefont{G.~A.} \bibnamefont{Sawatzky}},
  \bibinfo{journal}{Phys. Rev. B} \textbf{\bibinfo{volume}{41}},
  \bibinfo{pages}{928} (\bibinfo{year}{1990}).

\bibitem[{\citenamefont{Koster et~al.}(1998)\citenamefont{Koster, Kropman,
  Rijnders, Blank, and Rogolle}}]{Koster1998}
\bibinfo{author}{\bibfnamefont{G.}~\bibnamefont{Koster}},
  \bibinfo{author}{\bibfnamefont{B.~L.} \bibnamefont{Kropman}},
  \bibinfo{author}{\bibfnamefont{G.}~\bibnamefont{Rijnders}},
  \bibinfo{author}{\bibfnamefont{D.~H.~A.} \bibnamefont{Blank}},
  \bibnamefont{and} \bibinfo{author}{\bibfnamefont{H.}~\bibnamefont{Rogolle}},
  \bibinfo{journal}{Appl. Phys. Lett.} \textbf{\bibinfo{volume}{73}},
  \bibinfo{pages}{2920} (\bibinfo{year}{1998}).

\bibitem[{\citenamefont{Als-Nielsen and McMorrow}(2001)}]{als-nielsen2001}
\bibinfo{author}{\bibfnamefont{J.}~\bibnamefont{Als-Nielsen}} \bibnamefont{and}
  \bibinfo{author}{\bibfnamefont{D.}~\bibnamefont{McMorrow}},
  \emph{\bibinfo{title}{Elements of Modern X-ray Physics}}
  (\bibinfo{publisher}{Wiley}, \bibinfo{address}{New York},
  \bibinfo{year}{2001}).

\bibitem[{\citenamefont{L.Henke et~al.}(1993)\citenamefont{L.Henke, Gullikson,
  and Davis}}]{Henke1993}
\bibinfo{author}{\bibfnamefont{B.}~\bibnamefont{L.Henke}},
  \bibinfo{author}{\bibfnamefont{E.~M.} \bibnamefont{Gullikson}},
  \bibnamefont{and} \bibinfo{author}{\bibfnamefont{J.}~\bibnamefont{Davis}},
  \bibinfo{journal}{Atomic Data and Nuclear Data Tables}
  \textbf{\bibinfo{volume}{54}}, \bibinfo{pages}{181} (\bibinfo{year}{1993}).

\bibitem[{\citenamefont{Jackson}(1962)}]{Jackson}
\bibinfo{author}{\bibfnamefont{J.~D.} \bibnamefont{Jackson}},
  \emph{\bibinfo{title}{Classical electrodynamics}}
  (\bibinfo{publisher}{Wiley}, \bibinfo{address}{New York},
  \bibinfo{year}{1962}).

\bibitem[{\citenamefont{Soufli and Gullikson}(1997)}]{Soufli1997}
\bibinfo{author}{\bibfnamefont{R.}~\bibnamefont{Soufli}} \bibnamefont{and}
  \bibinfo{author}{\bibfnamefont{E.~M.} \bibnamefont{Gullikson}},
  \bibinfo{journal}{Appl. Opt.} \textbf{\bibinfo{volume}{36}},
  \bibinfo{pages}{5499} (\bibinfo{year}{1997}).

\bibitem[{\citenamefont{Alders et~al.}(1997)\citenamefont{Alders, Hibma,
  Sawatzky, Cheung, van Dorssen, Roper, Padmore, van~der Laan, Vogel, and
  Sacchi}}]{alders:97a}
\bibinfo{author}{\bibfnamefont{D.}~\bibnamefont{Alders}},
  \bibinfo{author}{\bibfnamefont{T.}~\bibnamefont{Hibma}},
  \bibinfo{author}{\bibfnamefont{G.~A.} \bibnamefont{Sawatzky}},
  \bibinfo{author}{\bibfnamefont{K.~C.} \bibnamefont{Cheung}},
  \bibinfo{author}{\bibfnamefont{G.~E.} \bibnamefont{van Dorssen}},
  \bibinfo{author}{\bibfnamefont{M.~D.} \bibnamefont{Roper}},
  \bibinfo{author}{\bibfnamefont{H.~A.} \bibnamefont{Padmore}},
  \bibinfo{author}{\bibfnamefont{G.}~\bibnamefont{van~der Laan}},
  \bibinfo{author}{\bibfnamefont{J.}~\bibnamefont{Vogel}}, \bibnamefont{and}
  \bibinfo{author}{\bibfnamefont{M.}~\bibnamefont{Sacchi}},
  \bibinfo{journal}{J. Appl. Phys.} \textbf{\bibinfo{volume}{82}},
  \bibinfo{pages}{3120} (\bibinfo{year}{1997}).

\bibitem[{\citenamefont{van~der Laan and Thole}(1988)}]{laan:88a}
\bibinfo{author}{\bibfnamefont{G.}~\bibnamefont{van~der Laan}}
  \bibnamefont{and} \bibinfo{author}{\bibfnamefont{B.~T.} \bibnamefont{Thole}},
  \bibinfo{journal}{Journ. Electr. Spectr. Relat. Phenom.}
  \textbf{\bibinfo{volume}{46}}, \bibinfo{pages}{123} (\bibinfo{year}{1988}).

\end{thebibliography}

\end{document}